\documentclass[aps,prl,twocolumn,showpacs,superscriptaddress]{revtex4-1}  
\usepackage{graphicx}  
\usepackage{dcolumn}   
\usepackage{bm}        
\usepackage{amssymb}   
\usepackage{hyperref}
\usepackage{color}
\usepackage{amsmath}
\usepackage{algorithm,algcompatible}

\hypersetup{
  colorlinks   = true, 
  urlcolor     = blue, 
  linkcolor    = black, 
  citecolor   = black 
}
\usepackage{lipsum} 
\usepackage{enumerate} 

\begin{document}

\newcommand{\todo}[1]{#1}
\newcommand{\removed}[1]{}


\title{Topological metric detects hidden order in disordered media}

\author{Dominic J. Skinner}
\affiliation{Department of Mathematics, Massachusetts Institute of Technology, Cambridge, Massachusetts 02139-4307, USA}
\author{Boya Song}
\affiliation{Department of Mathematics, Massachusetts Institute of Technology, Cambridge, Massachusetts 02139-4307, USA}
\author{\todo{Hannah Jeckel}}
\affiliation{Max Planck Institute for Terrestrial Microbiology, 35043 Marburg, Germany}
\affiliation{Department of Physics, Philipps-Universit\"at Marburg, 35043 Marburg, Germany}
\author{\todo{Eric Jelli}}
\affiliation{Max Planck Institute for Terrestrial Microbiology, 35043 Marburg, Germany}
\affiliation{Department of Physics, Philipps-Universit\"at Marburg, 35043 Marburg, Germany}
\author{\todo{Knut Drescher}}
\affiliation{Max Planck Institute for Terrestrial Microbiology, 35043 Marburg, Germany}
\affiliation{Department of Physics, Philipps-Universit\"at Marburg, 35043 Marburg, Germany}
\author{J\"{o}rn Dunkel}
\affiliation{Department of Mathematics, Massachusetts Institute of Technology, Cambridge, Massachusetts 02139-4307, USA}
\date{\today}

\begin{abstract}
Recent advances in microscopy techniques make it possible to study the growth, dynamics, 
and response of complex biophysical systems at single-cell resolution, from \todo{bacterial communities to 
tissues and organoids}. In contrast to ordered crystals, it is less obvious how one can 
reliably distinguish two amorphous yet structurally different cellular materials. Here, we introduce a topological earth mover's (TEM)  
distance between disordered structures that compares local graph neighborhoods
of the microscopic cell-centroid networks. \todo{Leveraging structural information contained in the neighborbood motif distributions, 
the TEM metric allows an interpretable reconstruction of equilibrium and non-equilibrium phase spaces and embedded pathways from static system snapshots alone. Applied to cell-resolution imaging data, 
the framework recovers time-ordering without prior knowledge about the underlying dynamics, revealing 
that fly wing development solves a topological optimal transport problem. Extending our topological 
analysis to bacterial swarms, we find a universal neighborhood size distribution 
consistent with a Tracy-Widom law.}
\end{abstract}

\maketitle


Discrete particulate objects, from atoms to cells, compose the majority
of physical and living systems. Modern microscopy and simulation techniques enable us to study the elementary building blocks of solids~\cite{RevModPhys.59.615,Sugimoto:2007aa}, 
colloidal and granular materials \cite{Glotzer:2007aa,Irvine,Nauer2019}, bacterial biofilms \cite{Raimo,Poon2018}, and
tissues \cite{DrosophilaExp} with unprecedented resolution over large scales.  These experimental and computational advances have highlighted the 
importance of local spatial organization~\cite{BiofilmFlow} and disorder~\cite{Goodrich:2014aa} for the global behaviors of both
equilibrium and non-equilibrium materials, spurring substantial theoretical efforts to link discrete microstructure with macroscale properties. 
Prime examples include the recent successful characterizations of epithelial cell (EC) 
layers  and other cellular materials  through the geometric~\cite{ManningPRX,ManningNaturePhysics} and topological \cite{LazarPRL,
JensenShannonPRE,CellComplexesPRE,Wenzel2019} analysis of Voronoi and Delaunay tesselations~\cite{VoronoiBook}. In spite of such major progress, 
 high-resolution data continue to pose fundamental conceptual and practical challenges regarding the proper classification of  \todo{discrete physical and biological structures. Specifically, it is still unclear whether one can recover parametric embeddings, phase space dimensions  and time ordering from a topological analysis of static snapshots alone, and whether such analysis can help reveal the governing principles of multicellular development.}

\par
\begin{figure*}[ht!]
\includegraphics{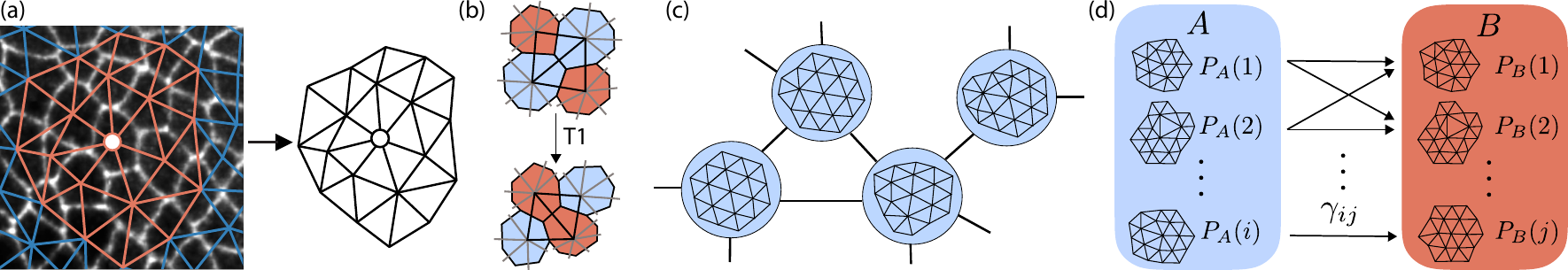}
\caption{\label{fig1}Key conceptual steps for calculating the topological earth mover's (TEM) distance illustrated for an epithelial cell layer. 
			(a)~Experimental image of epithelial cell layer from
			\emph{Drosophila} embryo (adapted with permission from Ref.~\cite{DrosophilaExp,DrosophilaNum}) with
			Delaunay triangulation overlayed (blue, red).
			The local neighborhood network of radius $r=2$ (red) 
			is shown for a selected  vertex (white circle).
			(b)~A \todo{topological} T1 transition, corresponding to an exchange of neighbors, is reflected in a 
			change of the local radius-2 neighborhood. Voronoi cells (red, blue) with Delaunay triangulation 
			overlayed (black, grey) are shown.
			(c)~Nodes of the flip graph correspond to networks, and 
			are connected by an edge if the networks are one
			T1 transition (or flip), away from each other.
			(d)~The map $\gamma_{ij}$ transports the distribution $A$ to
			the distribution~$B$.}
\end{figure*}

To tackle these problems, we introduce here a topological \todo{earth mover's (TEM)} distance by combining ideas from statistical topology~\cite{LazarPRL,JensenShannonPRE,CellComplexesPRE,HenrikLeaf,Katifori2019Topology} and optimal transport theory~\cite{SolomonTransport,solomon2016continuous} with non-equilibrium statistical mechanics\todo{~\cite{Ronhovde2012}}. The TEM metric compares two  discrete material structures by quantifying the statistical differences in the local
network topology of their Delaunay triangulations (Fig.~\ref{fig1}). Intuitively, computing TEM($A$, $B$) amounts to estimating the smallest number of edge-flips needed to make the local network topology of material $A$ statistically indistinguishable from the local network topology of material $B$. Physically, this procedure can be interpreted as finding the average lowest-energy path connecting two disordered structures, and we provide an efficient algorithm for realizing this computationally demanding task for systems with $\sim 10^6$ particles~\cite{SM}.

\par
To demonstrate the practical potential of this framework for the analysis of both equilibrium and non-equilibrium systems, 
we  \todo{present a broad set of } applications:
First, we show that the TEM  metric successfully distinguishes jammed disordered packings of both monodisperse 
and polydisperse ellipsoids. \todo{Thereafter, we use the TEM framework to  reconstruct the non-equilibrium phase diagram of active Brownian particle (ABP) simulations 
 without recourse to time resolved data. Next, by measuring the pairwise TEM distances between unsorted experimental images of a developing fruit 
fly wing, we are able to reconstruct their temporal ordering and discover that wing development follows an optimal
transport geodesic, suggesting a previously unrecognized optimization principle in tissue development. Finally, by extending our topological 
analysis to single-cell resolution data from bacterial swarming experiments, we observe a universal neighborhood size distribution 
consistent with a Tracy-Widom law.}

To define the TEM distance, we consider the specific example of a two-dimensional (2D) cell layer as shown in Fig.~\ref{fig1}(a), 
although all subsequent definitions generalize to arbitrary point sets in $\mathbb{R}^2$ or $\mathbb{R}^3$. 
Our starting point is the Delaunay triangulation~\cite{VoronoiBook} of the cell centroid positions as shown in Fig.~\ref{fig1}(a).  In practice, it is often sufficient to take the positions 
of the EC nuclei as vertices of the Delaunay network~\cite{Idema2018}. If two random realizations of 
such networks are generated by the same physical or biological process, they will have different vertex 
positions and topology, but their local statistical properties (local connectivity patterns, etc.) will 
be identical provided the networks are sufficiently large. This fact has been exploited previously 
to define entropic~\cite{JensenShannonPRE} and earth mover's distances between cell 
complexes~\cite{CellComplexesPRE}. Here, we extend these ideas to define a physically motivated 
topological metric that measures statistical differences in the local Delaunay triangulations around vertices. 
Specifically, we define for each vertex a local neighborhood of radius $r$, which consists of all the 
vertices that are not more than $r$ edges away from the central vertex (see red subgraph corresponding 
to $r=2$ in Fig.~\ref{fig1}a).  We found that \todo{$r=2$ suffices for many practical applications~\cite{SM}. 
Although $r$ can, in principle, be chosen 
arbitrarily large, TEM computations become expensive for a larger neighborhoods}; 
we therefore focus on the case $r=2$ from now on. 
The local neighborhoods 
of two vertices are of the same topological type $i$ if they are  graph-isomorphic. Counting 
the occurrences of the various neighborhood types $i$ across all vertices yields \todo{a probability distribution $P(i)\ge 0$ that  characterizes 
the topo-statistical state of the cell network.}
\par
To provide an intuitive physical motivation for the TEM metric, let us recall that the Delaunay network is invariant under infinitesimal perturbations and can change only through a topological T1 transition (Fig.~\ref{fig1}b). 
For EC  layers there is an energy barrier to T1 transitions~\cite{Bi2014}, and so the energy cost to transform from one neighborhood type to
another is directly related to the number of T1 transitions 
required. For other packed systems there typically exist similar energetic cost for changing neighbors through T1 transitions.  
Motivated by this, we can define the energetic distance between two neighborhoods as the 
minimum number of T1 transitions separating 
them. This mathematically well-defined metric \cite{Lawson1972,SM} induces naturally a secondary graph structure,  known as the
flip graph \cite{Bose2009},  where nodes correspond to neighborhood types $i$ and are linked 
with an edge if they are one T1 transition away from  each other (Fig.~\ref{fig1}c). 
The minimum path length between two nodes on the flip graph is  the 
smallest number of T1 transitions needed to move between the 
corresponding neighborhood types. Moreover, the distribution $P(i)$ of neighborhood types in the EC layer can now be viewed as a distribution  on 
the nodes $i$ of the flip graph (blue box  in Fig.~\ref{fig1}d). 
\par

\begin{figure*}[t]
\includegraphics{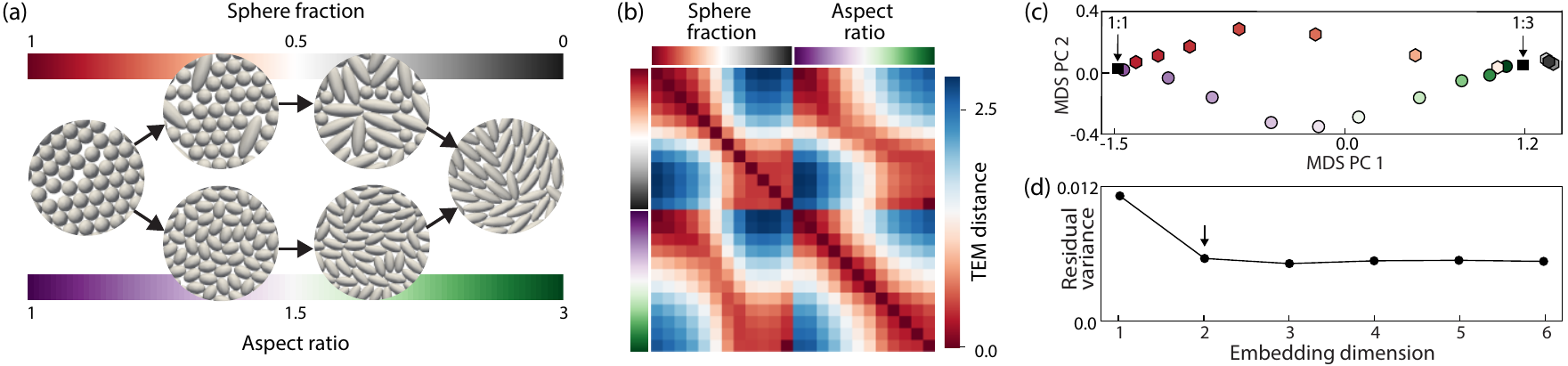}
\caption{\label{fig2}Poly- and monodisperse packings are distinguished by
		the TEM distance. 
		(a) Two alternative paths from aspect ratio 1:1 to aspect ratio 1:3 
		ellipsoid packings.
		Top is monodisperse with varying aspect 
		ratios, bottom is polydisperse with a mixture of 1:1 and 1:3 aspect
		ratios.
		(b) Distance matrix where each pixel represents the distance 
		between two simulated ellipsoid packings, each 
		containing 10,000 ellipsoids. 
		(c) Simulations are embedded in 2D using MDS, recovering two distinct 
		paths, one for monodisperse simulations (hexagons) and one for
		polydisperse simulations (circles).
		(d) The residual variance plateaus after embedding dimension 2, correctly identifying
		the true embedding dimension of the phase space (arrow).} 
\end{figure*}

Armed with this intuition, we can now define the TEM distance between the Delaunay triangulations of two materials $A$ and $B$ in a natural manner as 
the earth mover's or, equivalently, Wasserstein distance~\cite{SolomonTransport} between their neighborhood distributions $P_A$ and $P_B$ over the flip graph:  If $P_A(i)$ is the probability of neighborhood $i$ occuring in material
$A$, and $P_B(j)$ is the probability of neighborhood $j$ occuring in material
$B$, then a transport map, $\gamma_{ij}\ge 0$, from $A$ to $B$ satisfies
$\sum_j \gamma_{ij} =  P_A(i)$, $\sum_i \gamma_{ij} = P_B(i)$, see Fig.~\ref{fig1}(d). Then, the TEM distance between 
$A$ and $B$ is 
\begin{equation}
\label{e:TW}
\text{TEM}(A,B) = \min_{\gamma} \sum_{ij}  \gamma_{ij}\, d(i,j)
\end{equation}
where $d(i,j)$ is the distance between the neighborhoods $i$ and $j$ on the flip graph, and the minimum 
is taken over all possible transport maps $\gamma=(\gamma_{ij})$. 
We emphasize that, in contrast to widely used entropic distances measures 
between distributions~\cite{JensenShannonPRE}, the definition of TEM uses the physically relevant information encoded in the metric structure $d(i,j)$ of the underlying observable space, which in our case 
reflects the typical energy cost of a  T1 transitions between network motifs. As a consequence, TEM generally outperforms purely entropic Kullback-Leibler/Jensen-Shannon 
divergences when one needs to distinguish complex structures that are characterized by weakly overlapping distributions; see Ref.~\cite{SolomonTransportDiscrete} and 
the Supplemental Material~\cite{SM} for explicit examples.
\par
For large systems, the minimization problem~(\ref{e:TW})  becomes  
computationally challenging. We combined two algorithmic insights~\cite{SM} to calculate  
TEM efficiently for disordered materials with millions of particles.  Building on a modification 
of the Weinberg algorithm~\cite{WeinbergAlg},  our numerical scheme~\cite{SM} first determines   
the flip-graph distances $d(i,j)$ of $N$ observed neighborhood motifs in $O(N)$ steps.  Given $d(i,j)$,  
the minimization over the transport maps $\{\gamma_{ij}\}$ can be recast as an minimum cost flow 
problem~\cite{SM,SolomonTransport}, which is efficiently solved with linear programming \cite{JuMP}. To demonstrate the broad 
applicability of our TEM framework, we focus in the remainder on applications relevant to current major 
research areas: colloidal packings, collective far-from-equilibrium dynamics, tissue development, and \todo{spatio-temporally heterogeneous multicellular  systems}.

\begin{figure*}[t]
\includegraphics{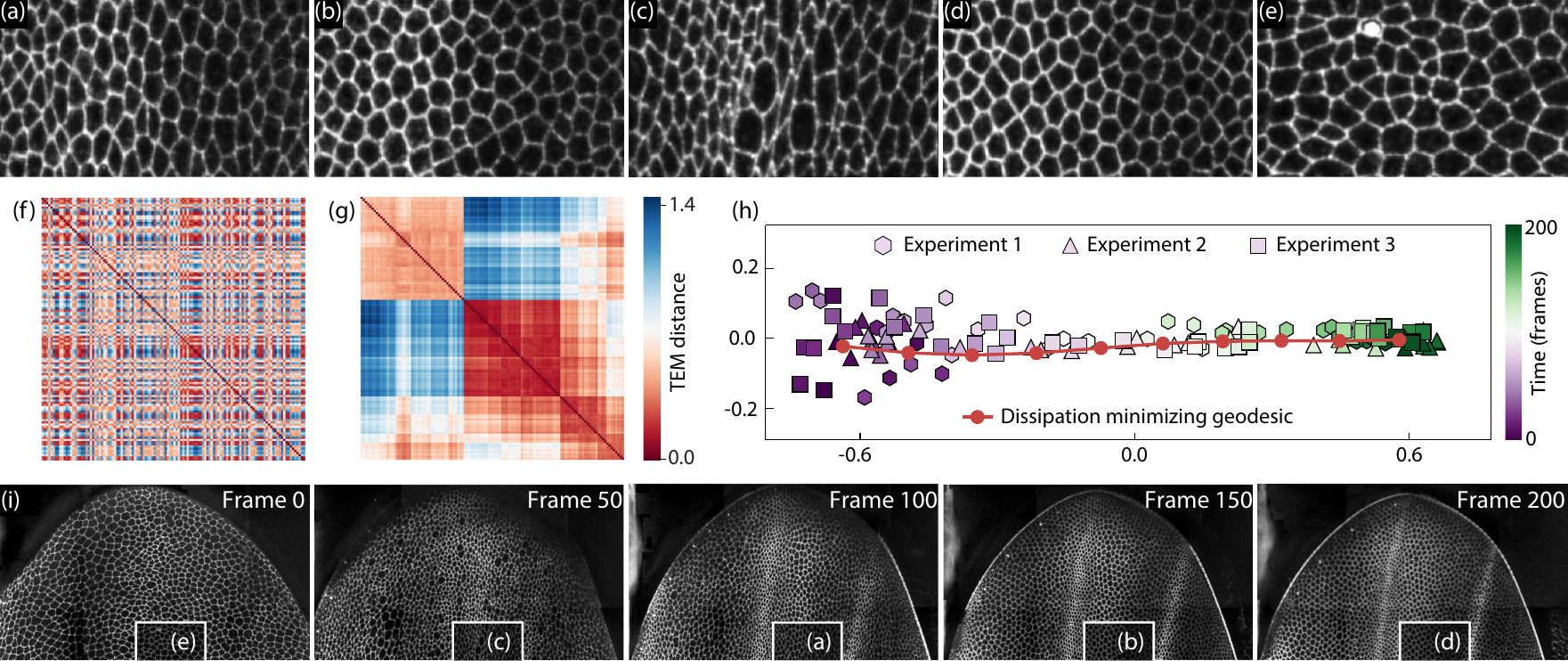}
\caption{\label{figDros}\todo{Developing \emph{Drosophila} embryos solve topological optimal transport problem, with MDS embedding
			recovering temporal order.}
			(a-e) Enlarged images showing epithelial cells at unknown times.
			Three experiments and 40 time frames per 
			experiment were used.
			(f) Matrix of TEM distances  between unsorted experimental images has no apparent structure.
			(g) TEM distance matrix sorted by hierarchical clustering  shows approximately three phases.
			(h) 2D MDS embedding recovers the temporal order as the principal component. \todo{Included in the embedding are
intermediate stages of a dissipation-minimizing~\cite{SM} geodesic between average start and end states, which the data falls on.}
			\removed{
			(i) Topological entropy also describes the principal component and decreases with time, indicative of non-equilibrium dynamics.}
			(i) The correct temporal ordering \todo{of the image time series} is recovered. The white boxes 
			show the source of images (a-e).  Data adapted with permission from \cite{DrosophilaExp,DrosophilaNum}.
}
\end{figure*}

\begin{figure*}[]
\includegraphics{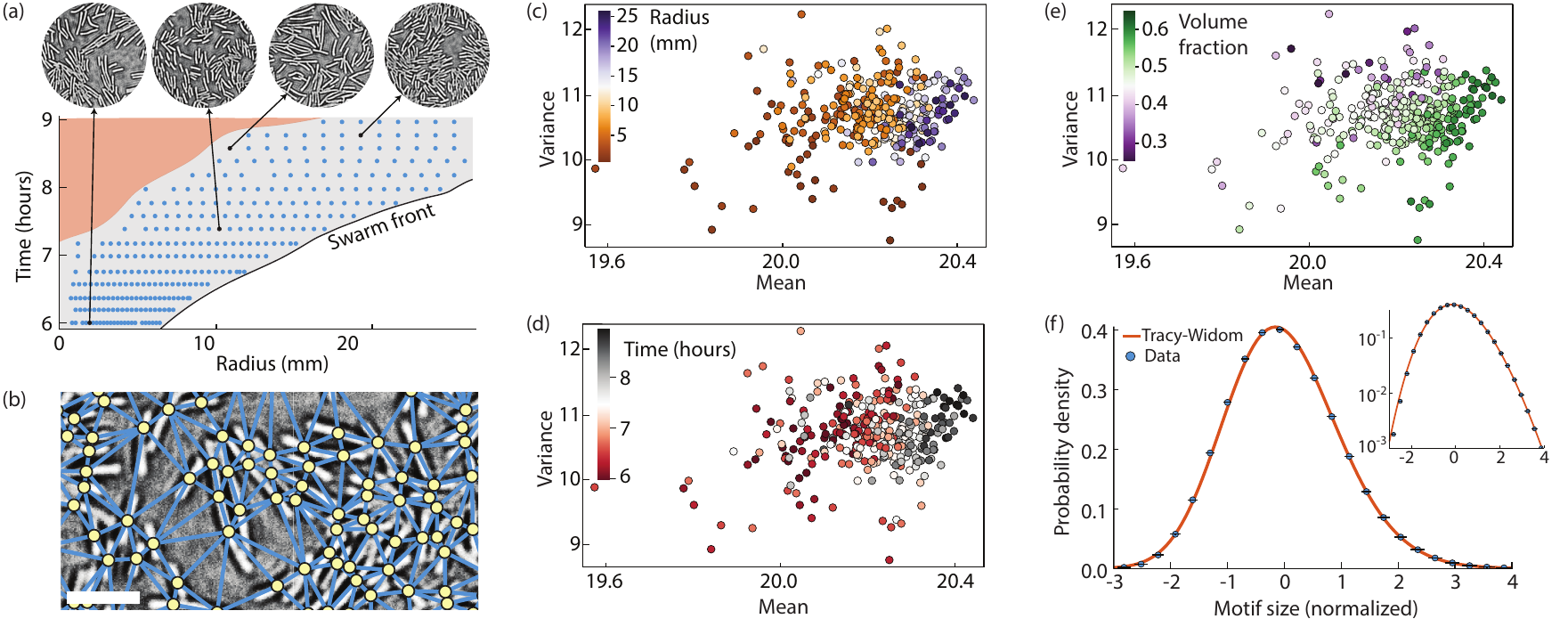}
\caption{\label{fig4}\todo{Motif sizes in heterogenous bacterial swarms exhibit universal non-equilibrium  statistics. (a) An azimuthally symmetric 
bacterial swarm expands as motile cells grow and divide~\cite{JeckelPNAS}.   We analyzed configurational 
snaphots at different space-time positions in the swarming monolayer phase (blue dots; select examples at top); the central multilayered biofilm phase (red) was excluded. 
(b) Aided by a machine learning algorithm, cell centroids (yellow circles) and their Delaunay tesselation  (blue lines) were determined. 
Scale bar: $10\mu \text{m}$. 
(c-e) The distribution of motif sizes, defined as the number of vertices per $r=2$ neighborhood, is heterogenous in space and time. 
The mean motif size increases, on average, with radial distance and time~(c,d). The variance decreases with area filling fraction (e), reflecting 
more ordered cell packings in dense swarming regions. 
(f) After normalizing to zero mean and unit variance, the combined histogram over all space-time snaphots (circles) follows a universal Tracy-Widom distribution (line). Horizontal error bars represent the standard deviation within a bin (inset: log scale plot).}
}
\end{figure*}
\par

Recent advances in the fabrication of geometrically complex colloids~\cite{Glotzer:2007aa,Irvine} and confocal imaging techniques~\cite{Raimo} 
have led to a renewed practical and theoretical interest in the characterization of granular~\cite{DonevMandM,Nauer2019} and 
biological materials~\cite{Raimo,Poon2018}.  Of particular importance in this context are the often fundamentally different behaviors of 
monodisperse~\cite{PackingRev}  and  polydisperse~\cite{TsimringGranularRev} colloidal systems. While the former are much better understood theoretically, 
the latter are often practically more relevant to natural systems and processes, such as particle segregation seen in industrial agriculture,
cereals, or avalanches~\cite{Kudrolli}. To demonstrate the usefulness of the TEM framework for capturing the \todo{essential topo-statistical}  differences between and across  
mono- and polydisperse systems, we generated jammed disordered packings of 10,000 ellipsoids using an event-driven packing code \cite{Donev2005}. Specifically, we were 
interested in distinguishing two different pathways for transitioning from a monodisperse packing of spheres (ellipsoids with aspect ratio 1:1) to a monodisperse packing of ellipsoids with aspect ratio 1:3 (Fig.~\ref{fig2}a). The first \lq monodisperse\rq{} transition path was realized by simulating 12 monodisperse packings of ellipsoids with aspect ratios varying from 1:1 to 1:3 (bottom path in Fig.~\ref{fig2}a). The second \lq polydisperse\rq{} transition path was realized by simulating  12  different binary mixtures of 1:1 and 1:3 ellipsoids (top path in Fig.~\ref{fig2}a). Computing the TEM distances between all $24\times 24$ pairs of simulations produces the symmetric TEM distance matrix shown in Fig.~\ref{fig2}(b). Given this matrix, it is natural to seek a faithful low-dimensional embedding in Euclidean space $\mathbb{R}^d$ that approximately preserves the TEM distance structure.
To construct the embedding we choose Multi-Dimensional Scaling (MDS), a generalized principal 
component (PC) analysis based on the TEM distance~\cite{MDS}. Since each pathway corresponds to a 
1D manifold (as only one parameter is varied in each case), the phase space
can be embedded in $\mathbb{R}^2$; indeed the $\mathbb{R}^2$ embedding clearly
distinguishes the two different pathways (Fig.~\ref{fig2}c). \todo{To find the 
dimensionality of the phase space, we calculate the
residual variance which plateaus at the relevant dimension~\cite{SM,Isomap} and correctly identifies} the ellipsoid 
embedding as 2D (Fig.~\ref{fig2}d).   \todo{We show in the Supplementary Information~\cite{SM} that the same approach can 
be used to infer the non-equilibrium phase space  of active Brownian particle (ABP) simulations \cite{Marchetti2014} from instantaneous system configurations.
More broadly, these examples illustrate how the TEM metric can discover phase spaces from 
configurational snapshots alone.}

\par
\todo{
In the remainder, we show that the topological analysis of data from two recent experiments~\cite{DrosophilaExp,JeckelPNAS} can reveal previously unrecognized biophysical optimization principles and universal statistical signatures. We begin by considering shuffled images (Fig.~\ref{figDros}a-e) of developing fruit fly embryo wings~\cite{DrosophilaExp,DrosophilaNum}. Using the Delaunay triangulation of the
cell centroids, hierarchical clustering \cite{HclustWard} of the TEM distance matrix of the shuffled images (Fig.~\ref{figDros}f) reveals 
three developmental main phases (Fig.~\ref{figDros}g). The resulting MDS embedding is essentially 1D, with the first principal component corresponding to 
time (Fig.~\ref{figDros}h), and thus restores the temporal order of the data (Fig.~\ref{figDros}i). This shows how the TEM framework can be used to infer temporal ordering from
ensemble measurements~\cite{Phil}.  More importantly, however, the TEM analysis reveals the developmental trajectory of the fly wing follows a topological geodesic,  
a continuous curve that minimizes the total length with respect to the TEM distance. Whilst earth mover's geodesics are in general not unique,  a unique path can be found by additionally minimizing transport dissipation~\cite{solomon2016continuous,SM}. The data fluctuates closely around this minimum-dissipation geodesic 
(red curve in Fig.~\ref{figDros}h), meaning that fly wing development approximately solves a dissipation-constrained topological optimal
transport problem.}

\todo{
Finally, we provide a more detailed characterization of the neighborhood motif distributions in 2D non-equilibrium systems. To this end, we analyze recent bacterial swarming experiments~\cite{JeckelPNAS} using machine learning~\cite{SM}  to identify individual cells (Fig.~\ref{fig4}a,b).  By determining the motif size distributions for snapshots taken at different  space-time locations in a growing swarm, we find that both mean and variance vary systematically with space, time, and 
cell density (Fig.~\ref{fig4}c-e). Strikingly, after rescaling to zero mean and unit variance, the combined motif size distribution closely matches~\cite{SM} a universal Tracy-Widom (TW) distribution (Fig.~\ref{fig4}f). TW distributions were recently reported for growing fluctuating fronts~\cite{Takeuchi2011}, dynamics of self-assembly~\cite{Makey2020}, active particle dynamics~\cite{Zeying2018,Tracy2009}, and phase transitions between strongly and weakly coupling regimes~\cite{Majumdar_2014}. We also find TW motif size distributions in the 
ABP and fly wing data when subsampling from the liquid-like phase~\cite{SM}, suggesting that TW distributions play a central role in the topo-statistics of non-equilbrium systems.
}

To conclude, the \todo{TEM metric framework} will be broadly applicable, from single-cell RNA-sequencing~\cite{SingleCellSeq}, cryo-electron microscopy~\cite{cryoEMFisher} and \todo{organoid characterization~\cite{Dekkers:2019aa}},
to structural  transitions in living~\cite{ManningPRX,Raimo} and nonliving~\cite{Glotzer:2007aa} matter.  In particular, it enables a 
direct comparison of the topological statistical properties of a wide range of fundamentally different systems, 
the only requirement being that transitions between basic motifs (Delaunay neighborhood structures, DNA strings, etc.)  can be mapped onto a joint flip-graph structure.

\par
This work was supported by a \todo{MathWorks Fellowship (D.J.S.)}, a James S. McDonnell Foundation Complex Systems Scholar Award (J.D.), and \todo{the Robert E. Collins Distinguished Scholar Fund (J.D.)}.
We thank \todo{Martin Abt} for help with cell segmentation and the MIT SuperCloud and Lincoln Laboratory Supercomputing Center for providing HPC resources.
%

\clearpage

\onecolumngrid
\hypersetup{
  colorlinks   = true, 
  urlcolor     = blue, 
  linkcolor    = black, 
  citecolor   = black 
}

\renewcommand{\d}{\text{d}}
\renewcommand{\div}[2]{\frac{\d {#1}}{\d {#2}}}
\renewcommand{\v}{\boldsymbol}
\setcounter{equation}{0}
\setcounter{figure}{0}
\setcounter{table}{0}
\setcounter{page}{1}
\renewcommand{\theequation}{S\arabic{equation}}
\renewcommand{\thefigure}{S\arabic{figure}}
\renewcommand{\bibnumfmt}[1]{[S#1]}
\renewcommand{\citenumfont}[1]{S#1}

\begin{center}
  \textbf{\large Supplemental Material:\\Topological metric detects  hidden order in disordered media}\\[.2cm]
Dominic J. Skinner,$^{1}$ Boya Song,$^{1}$, Hannah Jeckel$^{2,3}$, Eric Jelli$^{2,3}$, Knut Drescher$^{2,3}$, and J\"{o}rn Dunkel$^1$\\[.1cm]
 {\itshape ${}^1$Department of Mathematics, Massachusetts Institute of Technology, Cambridge, Massachusetts 02139-4307, USA\\}
 {\itshape ${}^2$Max Planck Institute for Terrestrial Microbiology, 35043 Marburg, Germany\\}
 {\itshape ${}^3$Department of Physics, Philipps-Universit\"at Marburg, 35043 Marburg, Germany\\}
(Dated: \today)\\[1cm]
\end{center}
  
\subsection{Voronoi diagrams and Delaunay triangulations}
Given a set of points $X$ in $\mathbb{R}^n$, the Voronoi tessellation partitions $\mathbb{R}^n$ 
into regions known as Voronoi cells. A point $y$ is in the Voronoi cell associated with
$x\in X$ if $y$ is closer to $x$ than to any other $z \in X \setminus \{x \} $. The Delaunay
triangulation is a graph with vertices at the points in $X$, with two points sharing an edge
if their corresponding Voronoi cells share a face. For more properties refer to~\cite{SMVoronoiBook}. 

\subsection{Interior and exterior points}
\begin{figure}[H]
	\centering
	\includegraphics{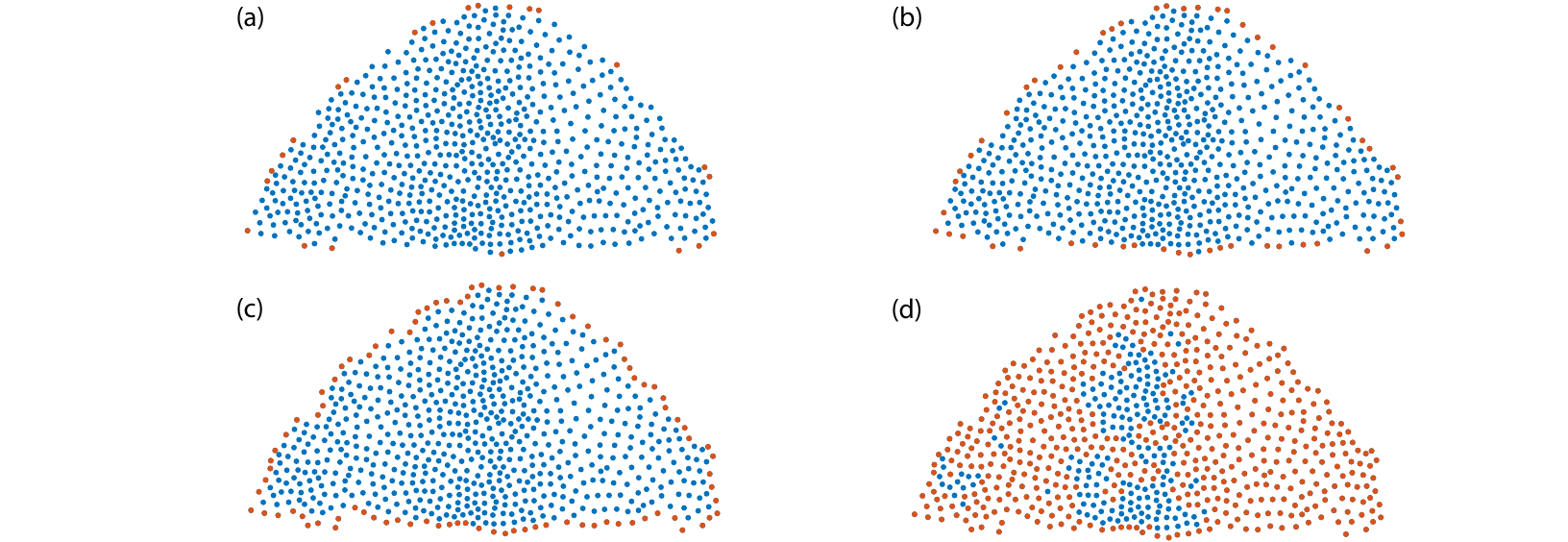}
	\caption{Taking the concave hull identifies exterior points for 
				segmented cell data from a \emph{Drosophila} 
				embryo~\cite{SMDrosophilaExp}. Cells centroids are shown in red 
				for exterior points, and blue for interior. (a) Exterior cells 
				are identified by the convex hull, which underidentifies 
				exterior points. (b) Edge cells are identified using
            $\alpha = 10\times$(median $\alpha_i$), underidentifying 
				exterior points. (c) $\alpha = 2\times$(median $\alpha_i$) 
				does a reasonable job at identifying exterior points.
            (d) $\alpha =$(median $\alpha_i$) overestimates exterior points.}
	\label{fig:alpha_shape_comp}
\end{figure}

For simulations with periodic boundary conditions the Deluanay graph extends periodically
and there are no boundary cases. For simulations or experiments that
are not periodic, there exist exterior points on the edge which may have quite different properties from 
points in the bulk. We do not wish to include edge effects and so 
we only take local networks of radius $r$ for points at least $r$ edges away from exterior points.
Exterior points are identified by calculating the alpha shape, or concave hull~\cite{SMalphaShape}.
In short, a point is an exterior point if a circle (or sphere in $\mathbb{R}^3$) of radius $\alpha$ can 
intersect that point without enclosing any other points. For every point $x_i$, there is a largest circle
which intersects $x_i$, but does not include any other points; call its radius $\alpha_i$. For the non-periodic
\emph{Drosophila} example, $\alpha = 2\times$(median $\alpha_i$) identifies cells
on the edge without incorrectly identifying interior points (Fig.~\ref{fig:alpha_shape_comp}).

\subsection{Storing and comparing networks}
Many thousands of topologically distinct networks were observed, requiring
fast methods to store and compare them. The approach taken here was
to replace each network with a vector of integers that uniquely represents its
topological type. Once represented as a vector, the networks can be stored as a
dictionary with an $O(\ln N)$ cost to read for $N$ topologically distinct graphs. 
Calculating the probability distribution for $N$ networks is then $O(N\ln N)$, 
rather than the $O(N^2)$ cost that would be required if one used an algorithm
that could only compare two graphs at a time.

\begin{figure}\centering
\includegraphics[width=0.9\textwidth]{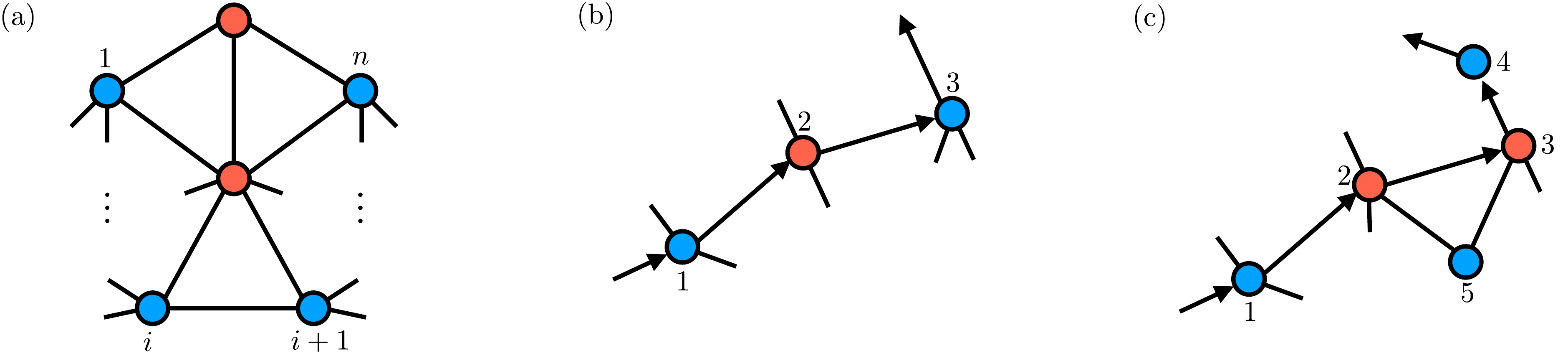}
\caption{\textbf{Claim:} The local network of radius $r$ formed from a Delaunay triangulation
	is triply connected. \\
	\textbf{Proof:} 
	First show the local network of radius $r=1$ is triply connected.
	Suppose two vertices are removed, if the central vertex is kept, as everything is connected to the
	central vertex, the graph is still connected. Suppose now that the central vertex and one other vertex
	are removed, as in (a). Label the remaining vertices by their anti-clockwise ordering about the central
	vertex for a particular embedding. Then as the local network is a triangulation, the central vertex and
	vertices $i$, $i+1$ form a triangle, so vertex $i$ is connected to $i+1$, hence the network 
	remains connected.\\
	Now consider a local network of arbitrary radius, and a path between two vertices, $A$ and $B$. 
	If no vertices on this path are removed $A$ and $B$ are still connected. If one vertex is removed, 
	vertex 2 in (b), the path enters the vertex from 1 and leaves toward 3. But 1 and 3 are in the local
	network of radius 1 around vertex 2, hence are connected, and so an alternative path from $A$ to $B$
	can be found. \\
	The final case is when vertex 3 is also deleted, so an alternative path to vertex 4 must be found, 
	as in (c). Here, note that as the network is a triangulation, vertices 2 and 3 are connected to a common
	vertex, vertex 5 which is in the local network of radius 1 for both. So 1 and 5 are connected as the
	local network of radius 1 is connected around 2, but 5 and 4 are also connected as the local network
	of radius 1 is connected around 3. Therefore an alternative path from $A$ to $B$ can be found, hence the
	network remains connected. }\label{fig:TriplyConnected}
\end{figure}

To encode the topology we use a modified Weinberg algorithm. The Weinberg
algorithm uniquely encodes the topology of a triply connected planar graph,
where triply connected means at least three vertices need to be removed to
disconnect the graph~\cite{SMWeinbergAlg,SMLazarPRL}. The local network is triply
connected, see Fig.~\ref{fig:TriplyConnected}.

In short, Weinberg's algorithm canonically labels Eulerian circuits, and from all possible
Eulerian circuit picks the lexographically first labeling. This labeling is taken as the 
vector; every isomorphic graph has the same vector and if two graphs have the same vector 
they are isomorphic.
The procedure for finding the Eulerian circuit from a given oriented starting edge is
detailed in algorithm \ref{Alg:Tremaux}, and the canonical labeling for a particular
starting edge is detailed in algorithm \ref{Alg:WeinbergV}. 
The total algorithm is described in algorithm \ref{Alg:WeinbergT}. 

\begin{algorithm}[H]
\caption{Tr\'emaux's algorithm for finding an Eulerian circuit in a directed graph with a particular
edge to be traversed first~\cite{SMWeinbergAlg}.}\label{Alg:Tremaux}
\begin{algorithmic}[1]
\State \textbf{Input:} A directed graph and a chosen edge.
\State The first vertex is the source of the chosen edge, the first step is to the destination of the
edge. No edge is traversed twice and future steps are made according to the following rules:
\State If a new vertex is reached, exit this vertex with the outgoing edge to the right
of the edge that you entered from.
\State If a previously visited vertex is reached, exit, if possible, towards the vertex that
  you were previously at.
\State If a previously visited vertex is reached, and it is not possible to exit towards
  the vertex that you were previously at, exit to the nearest available outgoing 
  edge to the right.
\State If there are no edges available, the algorithm terminates, and a Eulerian circuit has
    been found.
\end{algorithmic}
\end{algorithm}

\begin{algorithm}[H]
\caption{Canonical labeling for a graph using a specific
oriented edge to start. A worked example is shown in Fig.~\ref{fig:weinberg_demo}.}\label{Alg:WeinbergV}
\begin{algorithmic}[1]
\State \textbf{Input:} an undirected graph and an edge for which a direction is chosen.
\State Replace every edge of the graph with two directed edges oriented in opposite
       directions.
\State Take the Eulerian circuit starting with the chosen edge according to algorithm~\ref{Alg:Tremaux}.
\State Label the starting vertex as 1, and as the circuit is traversed, label every 
       new vertex reached with consecutive integers. 
\State The labeling vector for this starting edge
       is the ordered record of vertices that are seen as the circuit is traversed
       (so if a vertex is crossed $n$ times it appears in the vector $n$ times).
\end{algorithmic}
\end{algorithm}

\begin{algorithm}[H]
\caption{Algorithm for finding a graph's Weinberg vector (modified 
	from~\cite{SMWeinbergAlg})}\label{Alg:WeinbergT}
\begin{algorithmic}[1]
\State \textbf{Input:} A local network (an undirected graph with central vertex). 
\FOR{edges connected to central vertex}
\State Find canonical labeling as described in algorithm \ref{Alg:WeinbergV} with
	the edge oriented outwards.
\ENDFOR
\State Do the above for the mirrored embedding of the network as well. Collect all
	resulting labelings.
\State All of the labelings are vectors of integers with length $2|E|$. Lexographically
    sort these vectors and take the first sorted vector. This first vector is the
    Weinberg vector for the network. 
\end{algorithmic}
\end{algorithm}

Weinberg's algorithm as described in~\cite{SMWeinbergAlg}
differers from algorithm \ref{Alg:WeinbergT} in that it traverses every edge in both directions
rather than just the edges originating at the central vertex and so recognizes the same graph with
different embeddings as isomorphic. The local Delaunay network is a near triangulation meaning all of
the faces are triangular except the one at infinity~\cite{SMBose2009}, so there is only one embedding
that we will observe (together with the mirrored embedding). We also only define isomorphism between 
egocentric networks to mean the networks are isomorphic and they have the same central vertex 
(it is possible for an egocentric network to have two possible candidates for the central vertex,
although these are quite rare in practice). Therefore to check if two egocentric networks are isomorphic
we need only consider the labelings that start by moving away from the central vertex. If they
are isomorphic they share the same labelings. The resulting vector will not be the same as 
the vector calculated by trying all edges, but still works as a topological identifier when
compared only to other vectors calculated in the same manner.
\begin{figure}[H]
	\centering
	\includegraphics[width=0.3\textwidth]{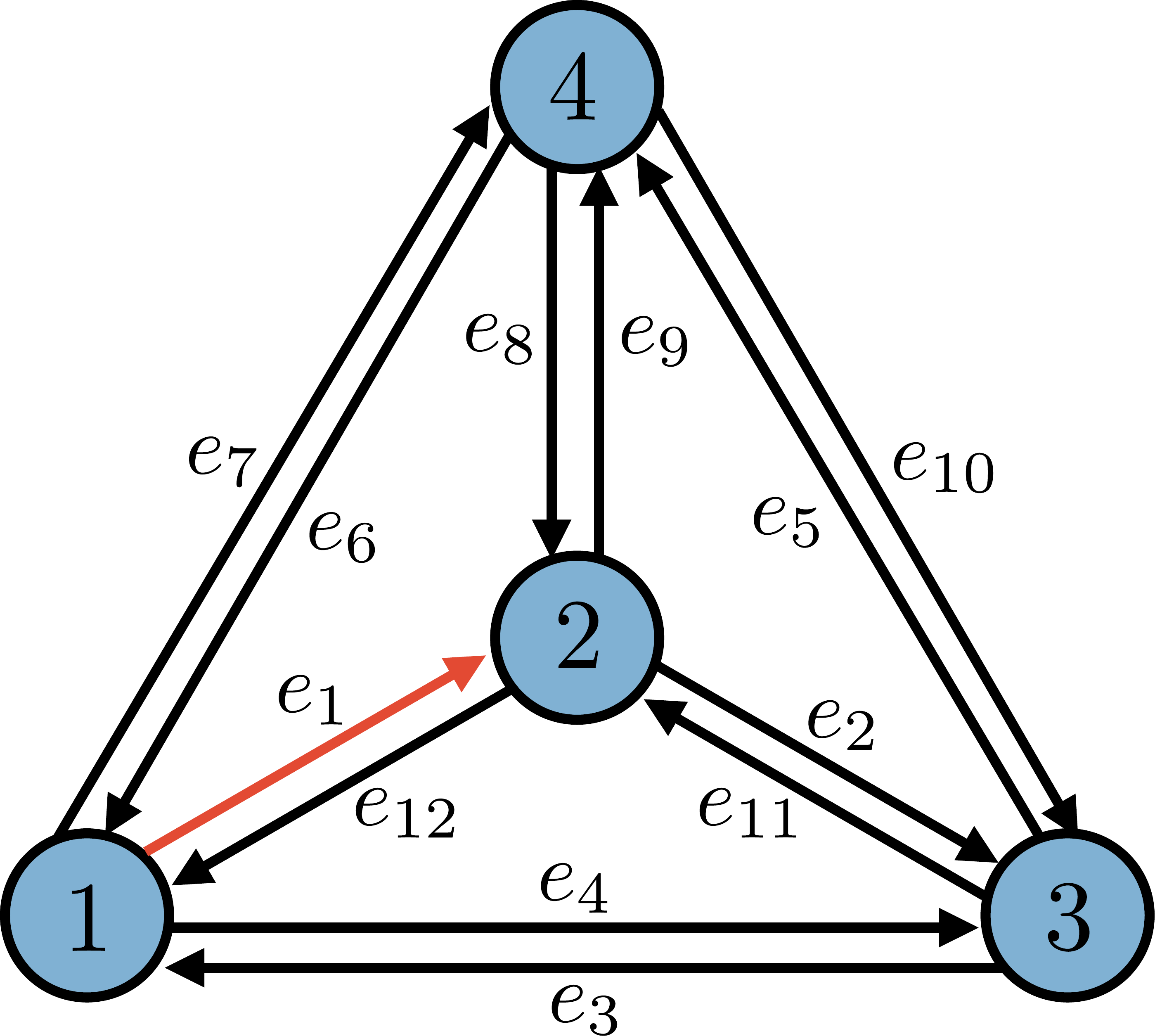}
	\caption{Calculation of a labeling vector from a specific starting edge for an
	    example local network of radius 1. The edge labeled $e_1$ is the starting edge,
	    and the first step is from vertex 1 to vertex 2.
            Next we exit to the right along edge $e_2$ to vertex 3. We have reached a new
            vertex so we exit to the right along edge $e_3$. Vertex 1 has been visited before
            so we take edge $e_4$ back to vertex 3. Vertex 3 has been visited before, but we
            have already used edge $e_3$, so take the nearest available edge to the right,
            edge $e_5$, onto vertex 4. We continue as before, using algorithm 
	    \ref{Alg:WeinbergV}, until we run out of edges. The edges are labeled
            in the order in which they are used. The canonical labeling is the record of
	    which vertices are visited, which is (1,2,3,1,3,4,1,4,2,4,3,2,1). 
           }
	\label{fig:weinberg_demo}
\end{figure}

The algorithm in~\cite{SMWeinbergAlg} finds the Weinberg vector in $O(E^2)$ time, as
finding the path and labeling it (which can be done simultaneously) takes $O(E)$ 
time, and this must be done for $2|E|$ edges. By only taking the edges that start
at the central vertex, which is typically $O(6)$, the calculation grows like
$O(E)$ as the size of the local network grows. 
For local networks of radius $r=2$ this typically allows the algorithm to
run 10-20 times faster.

\subsection{Calculating the flip graph}
\begin{figure}\centering
\includegraphics{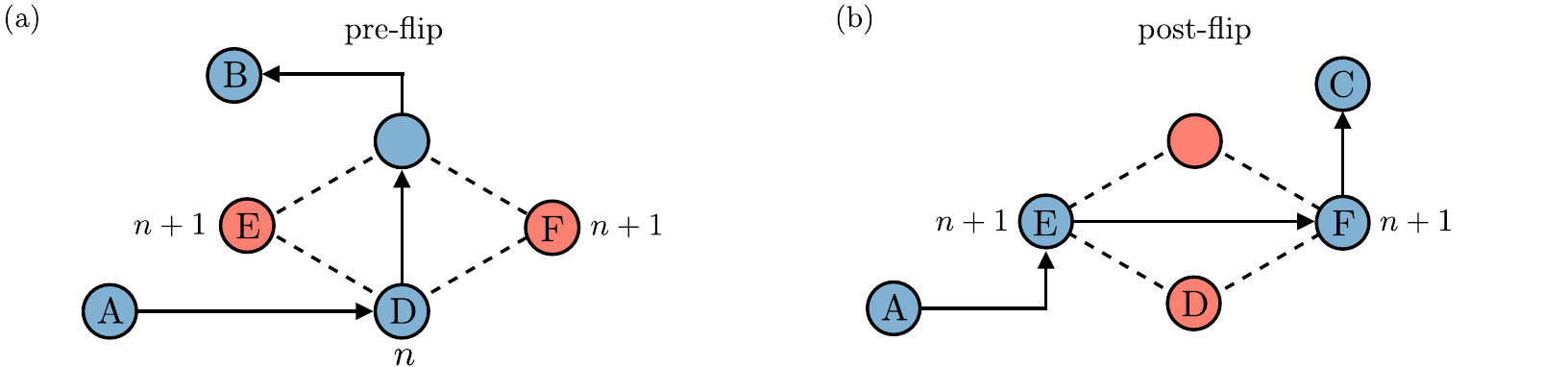}
\caption{\textbf{Claim:} A topological transition, or flip, either does not 
			increase all minimum path lengths from a specific vertex, or it 
			does not decrease all minimum path lengths. That is to say, given a 
			vertex, there cannot exist a vertex that gets closer after flipping 
			and a vertex that gets further away. 
			\textbf{Proof:} Suppose $A$ is our chosen vertex, $B$ is a vertex 
			which gets further away after flipping, and $C$ is a vertex which 
			gets closer after flipping. Therefore the pre-flip minimum path 
			from $A$ to $B$ goes through the edge that will be flipped, and 
			there are no other paths from $A$ to $B$ of the same length. In 
			particular, if $n$ is the distance between $D$ and $A$, then 
			$n+1$ must be the path distance between $E,F$, and $A$, else 
			flipping would not increase the distance. However, this means that 
			no post-flip minimum path from $A$ to $C$ can go through the edge 
			connecting $E$ to $F$ as they are both the same path length from 
			$A$. This means that the flip does not affect the path length
			from $A$ to $C$, a contradiction which proves the claim. }\label{fig:fliplength}
\end{figure}

\begin{figure}\centering
\includegraphics{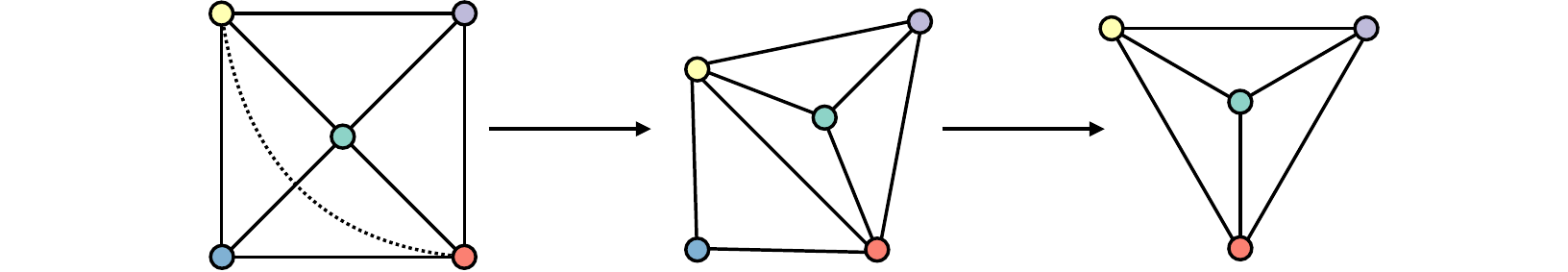}
\caption{Procedure for calculating the flip graph. 
		   (a) A local network of radius $r=1$, with a potential flip 
			identified (dotted line). 
			(b) The flip is performed, but the resulting graph is not a 
			local network. However, all path lengths from the central 
			vertex have not decreased, meaning the new local network of 
			radius 1 is a subgraph of this graph.
			(c) Taking the local network of radius 1 around the central 
			vertex (in green), gives a local network 1 flip away from 
			the original network.}\label{fig:FlipProcedure}
\end{figure}

Given $N$ observed networks, we wish to calculate which are connected by an edge in the 
flip graph, from which the minimum path length between two networks gives a measure of 
distance between them. This measure of distance is similar to the Levenshtein distance
between two strings, although in that case distances are calculated, as needed, using
dynamic programming~\cite{SMWagner}. Here, path lengths on the flip graph give distances, but this requires checking
if up to $N(N-1)/2$ edges exist. Instead we are able to calculate the flip graph
in $O(N E^2)$ time using algorithm \ref{Alg:FlipGraph}. The central insight is that after 
a T1 transition or flip, path distances to the central vertex either do not increase or do not decrease, as proved in
Fig.~\ref{fig:fliplength}. After flipping, the new graph need not be a local network, but due to 
the claim in Fig.~\ref{fig:fliplength}, for one of the local networks, flipping means the other 
local network is a subgraph of the flipped graph (see Fig.~\ref{fig:FlipProcedure}). Since this
requires calculating a Weinberg vector for potentially each edge of a network, the total cost
is $O(NE^2)$.

The flip graph is connected, but in practice this calculation often yields a disconnected graph. This occurs when
the path between two states goes through states that were not observed in the $N$ observed networks (but
do theoretically exist). If this occurs for a few isolated states that make up a negligible 
proportion of the total, then the largest
connected component can be taken and these isolated states can be ignored. If the number of
disconnected states is large, then the $N$ observed networks can be augmented by additional networks
observed in a Poisson-Voronoi process or similar~\cite{SMLazarPRL}, and a larger flip graph can be calculated. Networks
are added until the connected component of the new flip graph contains all (or almost all) of the original $N$ 
observed networks.

\begin{algorithm}[H]
\caption{Calculating the flip graph for $N$ networks}\label{Alg:FlipGraph}
\begin{algorithmic}[1]
\State \textbf{Input:} $N$ topologically distinct local networks each with central vertex known.
\FOR{1:N networks}
\FOR {edges in network}
\State Check if edge can be feasibly flipped~\cite{SMBose2009}. If it can check to see if the distances to the central
       vertex do not 
	\hspace*{1.8em} decrease (From claim in Fig.~\ref{fig:fliplength} finding at least one distance that
	increases or decreases suffices, and this can 
	\hspace*{1.8em} be done in the neighborhood of the flip).
\State If the distances do not decrease, find the new local network around the central vertex, which will
       be a 
	\hspace*{1.8em} subgraph of the flipped network. 
\State Connect the vertices corresponding to the original local network and the new local network in the flip
       graph.
\ENDFOR
\ENDFOR
\end{algorithmic}
\end{algorithm}

\subsection{Calculating the TEM distance}
The topological earth mover's (TEM) distance is defined to be the 
earth mover's distance between two probability distributions on the flip 
graph, 
\begin{subequations}
\begin{equation}
\text{TEM}(A,B) = \min_{\gamma} \sum_{i,j} \gamma_{ij} \; d(i,j),
\end{equation}
where the sum is taken over all pairs of networks $i,j$, $d(i,j)$ is the minimum
path length on the flip graph between networks $i$ and $j$, and $\gamma$ is
a map between distributions satisfying
\begin{equation}
\gamma_{ij} \geq 0, 
\qquad\qquad
\sum_j \gamma_{ij} = P_A(i), 
\qquad\qquad
\sum_i \gamma_{ij} = P_B(j). 
\end{equation}
\end{subequations}
Rather than optimize over all possible maps $\gamma$, the problem can be rephrased as a minimum
cost flow problem over the flip graph~\cite{SMSolomonTransport}. This is done by first converting 
to a minimum cost flow problem over the complete graph on $N$ vertices, where the weight, or
cost, on the edge between $i$ and $j$ is $d(i,j)$, and each vertex is a source or sink with 
strength $P_A(i) - P_B(i)$. This is then equivalent to solving on the flip graph, because the
cost of sending mass directly from $i$ to $j$ is the same cost as sending it through the minimum length
path between $i$ and $j$.

The minimum cost flow is converted into the standard formulation, 
by taking two additional nodes, one a source,
one a sink, and connecting every existing source to the new source 
by an edge with a capacity of the
existing source strength and connecting every existing sink to the new sink by an edge with demand of 
the existing sink strength. With the exeption of the new source and sink, all other sources and sinks are
then set to strength 0. To reduce the stiffness of the problem, each capacity was multiplied by
$1 + 10^{-5}$ and each demand was multiplied by $1-10^{-5}$. Under this relaxation of the problem,
the algorithm always converged, and the additional error was found to be $\sim 0.005\%$. For
significantly larger or stiffer problems, approximations to optimal transport, such as 
entropic regularization are possible~\cite{SMSolomonTransport}, but were not required here.

\subsection{Residual variance}
\begin{figure}
\includegraphics{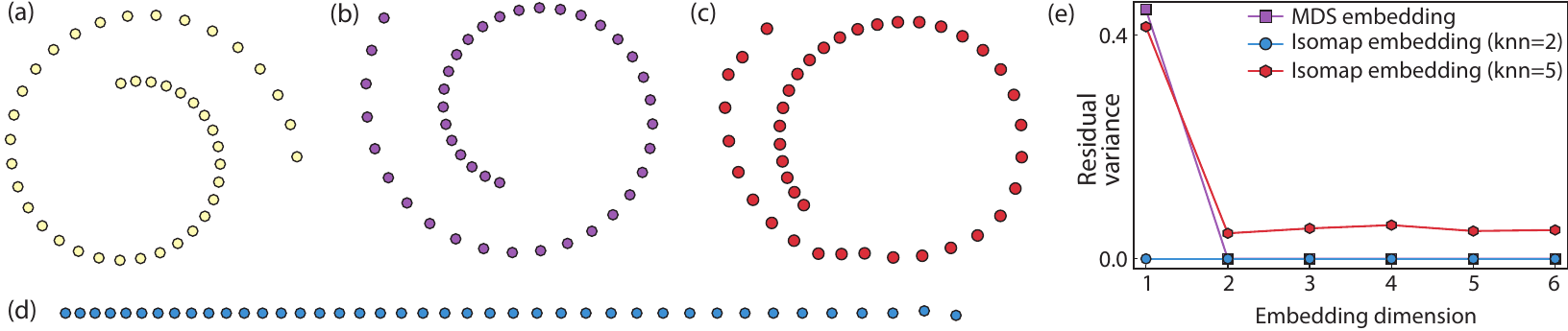}
\caption{\label{fig:IsomapMDS}Using Isomap to calculate residual variances recovers the correct manifold
	dimension, but an MDS embedding preserves the structure. (a) Points from a 2D spiral
	lie on a 1D manifold but have a non-trivial embedding in 2D space. (b) 2D MDS embedding (with Euclidean distances) of
	the data from (a) exactly preserves the structure. (c) 2D Isomap embedding (knn = 5)
	mostly preserves the structure. (d) 2D Isomap embedding (knn = 2)
	is effectively 1D. (e) The MDS embedding has no residual variance for dimension $\geq 2$ suggesting
	that the data is 2D, whereas the Isomap embedding with knn = 2 has almost no residual
	variance for a 1D embedding, correctly identifying that the data lies on a 1D manifold. 
	Isomap with knn = 5 suggests the data lies on a 2D manifold.}
\end{figure}
\begin{figure}
\includegraphics{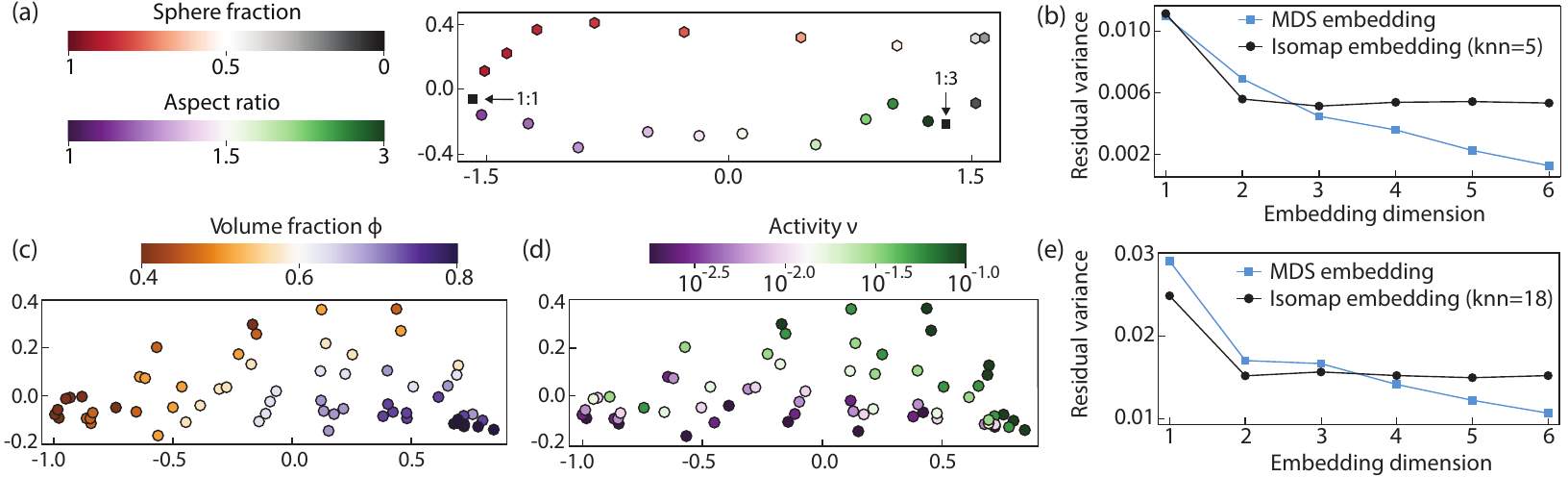}
\caption{\label{fig:ResidualVariance}Calculating the residual variance with Isomap finds 
	the dimension of the phase space, with the Isomap embedding comparable to the MDS embedding.
	(a) Isomap embedding of the ellipsoid packings with knn = 5 recovers the
	two distinct paths from 1:1 to 1:3 ellipsoid packings even more clearly than the MDS embedding (Fig.~2 main text).
	(b) The residual variance from the Isomap embedding plateaus after two dimensions indicating a 2D phase space,
	whereas the MDS residual variance does not clearly indicate any dimension. 
	(c-d) Isomap embedding of the ABP simulations with knn = 18 colored by volume fraction (c), 
	and activity (d). Both volume fraction and activity are recovered as coordinates,
	but do not correlate as strongly with the principal components of the embedding as the MDS embedding (Fig.~3
	main text).
	(e) Both in the MDS and Isomap embedding there is somewhat of a plateau after two dimension, but only Isomap
	definitively shows that the underlying space is 2D.} 
\end{figure}

The residual variance is defined as $1 - R^2(\hat{D},D_U)$, where $\hat{D}$ is the (Euclidean) 
distance matrix in the embedded space, $D_U$ is an unembedded distance matrix, and $R^2$ is the
linear correlation coefficient~\cite{SMIsomap}. To calculate the residual variance, we do not take
the TEM distance matrix, instead we take $D_U$ as the Isomap
distance matrix derived from the TEM distance matrix~\cite{SMIsomap}. In short, this means replacing the
TEM distance between two points with their distance along the manifold; local distances are preserved
and global distances become a sum of local distances along the path between two points~\cite{SMIsomap}.
Local here means the $k$ nearest neighbors of a point (knn), where $k$ is a parameter to be chosen.
The dimension of the manifold on which the points lie is the dimension for which the residual variance 
becomes negligible or does not decrease for higher dimensional embeddings~\cite{SMIsomap}.

To understand why Isomap produces the most accurate estimate of dimension,
consider the example of a 2D spiral; a 1D manifold with non-trivial embedding in 2D Euclidean
space (Fig.~\ref{fig:IsomapMDS}a). The 2D MDS embedding maintains the spiral (Fig.~\ref{fig:IsomapMDS}b),
as does a 2D Isomap embedding with a large number of neighbors (Fig.~\ref{fig:IsomapMDS}c).
In contrast, a 2D Isomap embedding using 2 neighbors ``unrolls'' the shape to get a straight line (Fig.~\ref{fig:IsomapMDS}d). 
Therefore, Isomap can correctly identify the manifold as 1D, whereas MDS incorrectly identifies the manifold
as 2D (Fig.~\ref{fig:IsomapMDS}e). That said, the MDS embedding preserves the non-trivial structure
of the manifold in 2D; for this reason we stick with MDS embeddings for visualization purposes.

For the ellipsoid packing and ABP examples, using the Isomap distance matrix to calculate the residual
variance correctly recovers the dimension of the subspaces, unlike MDS (Fig.~\ref{fig:ResidualVariance}b,e).
The number of neighbors used was chosen to be large enough to make the Isomap embedding 
consistent with the MDS embedding, whilst being small enough to ``unroll'' the manifold (Fig.~\ref{fig:ResidualVariance}a,c,d).

\subsection{Active Brownian particles}

Simulations of 2D active Brownian particles (ABPs) were performed with 2,000 particles following the method
described in~\cite{SMMarchetti2014}, which we briefly outline here. Periodic boundary conditions were used for a box of size 
$L\times L$. Let $\v{r}_i$ and $\theta_i$ describe the center position and the orientation 
of the $i$-th particle respectively. The over-damped dynamics of each particle is governed by the 
following equations,
\begin{align}
\div{\v{r}_i}{t} &= v \hat{\v{n}}_i + \mu \sum_{j\neq i}\v{F}_{ij}, \label{eq:ABP_r}\\
\div{\theta_i}{t} &= \eta_i(t),\label{eq:ABP_theta}
\end{align}

where $\hat{\v{n}}_i = [\cos\theta_i, \sin\theta_i]$ describes the orientation of the $i$-th particle, 
$v$ is the self-propulsion speed, and $\mu$ is the mobility. $\v{F}_{ij} $ 
is a pairwise soft repulsive force such that $\v{F}_{ij} =\v{0}$ when the particles $i$ and $j$ are 
not overlapping, and $\v{F}_{ij} = k(a_i+a_j-r_{ij})\hat{\v{r}}_{ij}$ with $r_{ij} = ||\v{r}_i - \v{r}_j||$ 
and $\hat{\v{r}}_{ij} = (\v{r}_i - \v{r}_j)/r_{ij}$ when the particles overlap. 
Eq. (\ref{eq:ABP_theta}) is a stochastic differential equation with Gaussian white noise, $\eta_i(t)$,
satisfying $\langle\eta_i(t)\eta_j(t')\rangle =2v_r\delta_{ij}\delta(t-t') $, 
where $v_r$ is the the rotational diffusion rate. The radius of $i$-th particle 
$a_i$ is drawn from a uniform distribution between $0.8a$ and $1.2a$. 
The domain size $L$ is computed from the volume fraction $\phi$ and the radii of the particles 
$L = \sqrt{\sum_i\pi a_i^2 /\phi}$. Choosing mean particle radius $a$ as the unit of length and the 
elastic time scale $\tau= (\mu k)^{-1}$ as the unit of time, the parameter space is reduced to the 
effective self-propulsion speed $\tilde{v}=v/(a\mu k)$, the packing fraction $\phi$ and the effective 
rotational diffusion $\tilde{v}_r=v_r/(\mu k)$. For all simulations, 
we fix $\tilde{v}_r = 5\times 10^{-4}$ and simulate values in
$\tilde{v}\in[1.8\times10^{-3}, 0.1]$ and $\phi\in[0.4,  0.8]$.

\todo{A custom, parallelized code employing graphics processing units (GPUs) was implemented to 
perform the simulations, following~\cite{SMMarchetti2014}}. We use a standard explicit Euler scheme to numerically integrate the 
dimensionless form of Eq. (\ref{eq:ABP_r}) and (\ref{eq:ABP_theta}) from $t=0$ to $t=5000\tau$ 
with a time step $\Delta t = 0.01\tau$. Only the snapshot of $t=5000\tau$ is used to calculate the
TEM distances, no dynamic information is used.

Following~\cite{SMMarchetti2014}, we neglect translational noise, although activity becomes equivalent
to translational noise in the limit where the orientational correlation time becomes much smaller
than the mean free time between collisions. For a given volume fraction, this limit will 
be realized as $v\to 0$, so reducing $v$ makes the system closer to a thermal system~\cite{SMMarchetti2014}.
That said, phase separation for active particle systems is a distinctly non-equilibrium phenomenon, 
making much of our simulated phase space far from equilibrium~\cite{SMMarchetti2014}.

\begin{figure}[t]
\includegraphics{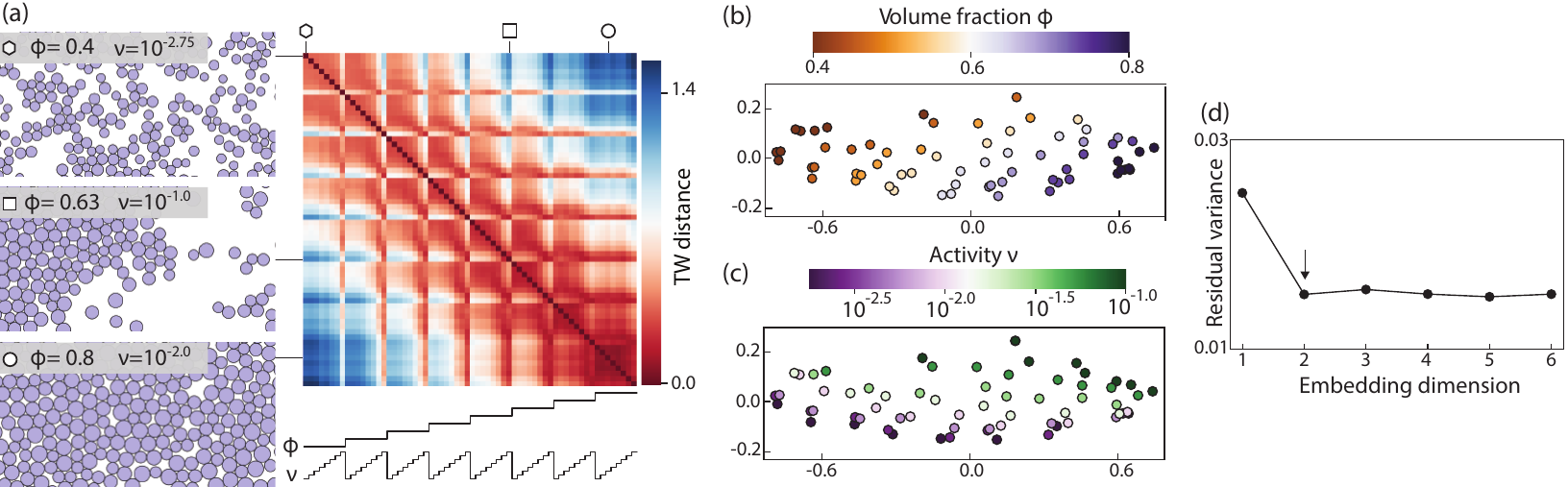}
\caption{\label{fig:ABP}\todo{Recovery of phase space dimension and coordinates
			for active Brownian particles.
			(a) Snapshots of simulations for different parameter values
			showing liquid like, phase separated, and glass like states, 
			and their corresponding entries in the distance matrix. For every 
			simulated parameter set, 5 runs with 2000
			particles were combined to obtain an average 
			distribution. 
			(b-d) 2D MDS embedding, (b-c) colored according 
			to volume fraction and activity respectively, showing the 
			original phase space is recovered.
			(d) The residual variance plateaus after embedding dimension 2, correctly identifying
			the true embedding dimension of the phase space (shown with arrow).}}
\end{figure}

\todo{Three representative partial simulation snapshots can be seen in Fig.~\ref{fig:ABP}(a). 
We computed the Delaunay tessellations of the ABP-centroid positions from each snapshot and then the associated $r=2$ 
motif distributions for each simulation $(\phi,v)$. The TEM distance matrix for all simulation pairs is depicted in 
Fig.~\ref{fig:ABP}(a). Notably, the first two principal components of the associated MDS embedding recover the phase space 
spanned by volume fraction and activity parameter (Fig. \ref{fig:ABP}b,c). 
In particular, the second principal component correlates closely with activity, demonstrating that the TEM 
metric detects the transition to far-from-equilibrium  dynamics (large $v$), which is recovered by the embedding 
without need for time-resolved data (Fig. \ref{fig:ABP}c). }
\subsection{Comparison with Jensen-Shannon}
\begin{figure}
\includegraphics[width=\textwidth]{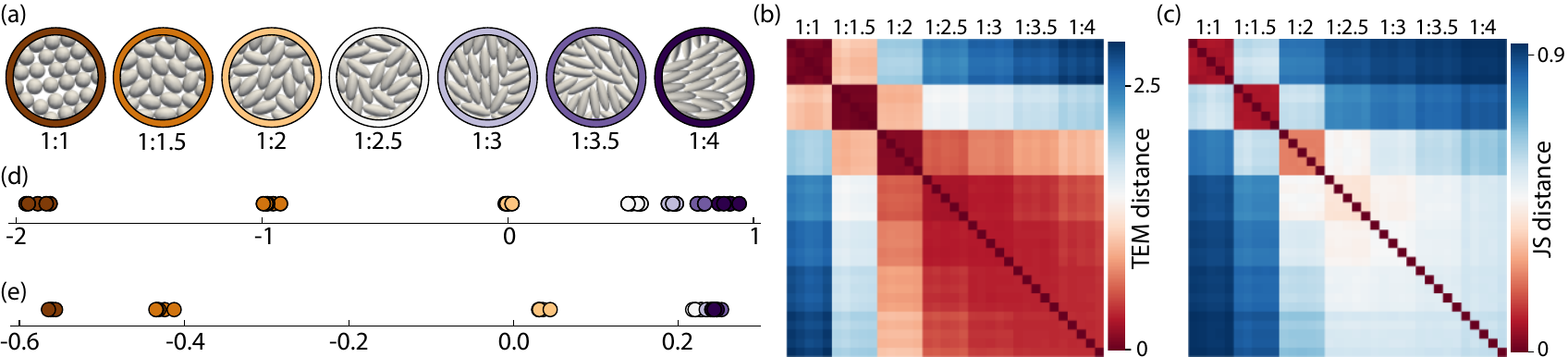}
\caption{\label{fig:JSComp} Comparison of TEM and JS distances ability to distinguish simulations of jammed disordered 
	packings of ellipsoids for different aspect ratios. 
	(a) Snapshots of ellipsoid packings for various 
	aspect ratios. 
	(b-c) Distance matrices for the TEM and JS distances respectively.
	Each pixel represents the distance between two simulations, with five 
	simulations	of 10,000 ellipsoids performed for each aspect ratio. 
	(d-e) 1D embedding of the simulations using
	MDS for TEM and JS respectively. Color coding is according
	aspect ratio as in (a). }
\end{figure}
The Wasserstein or earth mover's distance is only one of many possible
metrics that could be taken between two distributions. Another possible
distance is the Jensen-Shannon (JS) distance which has also been used to
distinguish cellular structures~\cite{SMJensenShannonPRE}. It is defined by
\begin{equation}
JS(A,B)^2 = \frac{1}{\ln 2} \sum_i \; \frac{1}{2} P_i^A \log \left[ \frac{P_i^A}{\frac{1}{2}(P_i^A+P_i^B)}\right] + 
\frac{1}{2} P_i^B \log \left[\frac{P_i^B}{\frac{1}{2} (P_i^A+P_i^B)} \right],
\end{equation}
where the sum is taken over all networks $i$, and $P_i^A$ is the probability of observing
network $i$ in distribution $A$, similarly for $P_i^B$. The JS distance is an entropic
distance between distributions, based on the idea of mutual information. 
It does not use any notion of distance between networks, only using their isomorphism 
classification. This has the drawback that while it can distinguish distinct distributions,
it cannot tell to what degree they are different, for example all non-overlapping distributions
are JS distance 1 away from eachother regardless of their particular forms~\cite{SMSolomonTransportDiscrete}.

For monodisperse packings with varying aspect ratio (Fig.~\ref{fig:JSComp}a),
both the TEM distance, and the JS distance are consistent across different
simulations with the same parameters (Fig.~\ref{fig:JSComp}b,c). Unlike the 
JS distance, the
TEM distance recognizes different simulations of 1:3-4 ellipsoids as
very similar to each other. The distance matrix was embedded
in a 1D space, the true dimension of the manifold on which
the data lies, using MDS. 
For the TEM distance this recovers the correct ordering of 
aspect ratios, whereas the JS distance is unable to separate some 
of the larger aspect ratio packings, Fig.~\ref{fig:JSComp}d,e.

\begin{figure}
\includegraphics{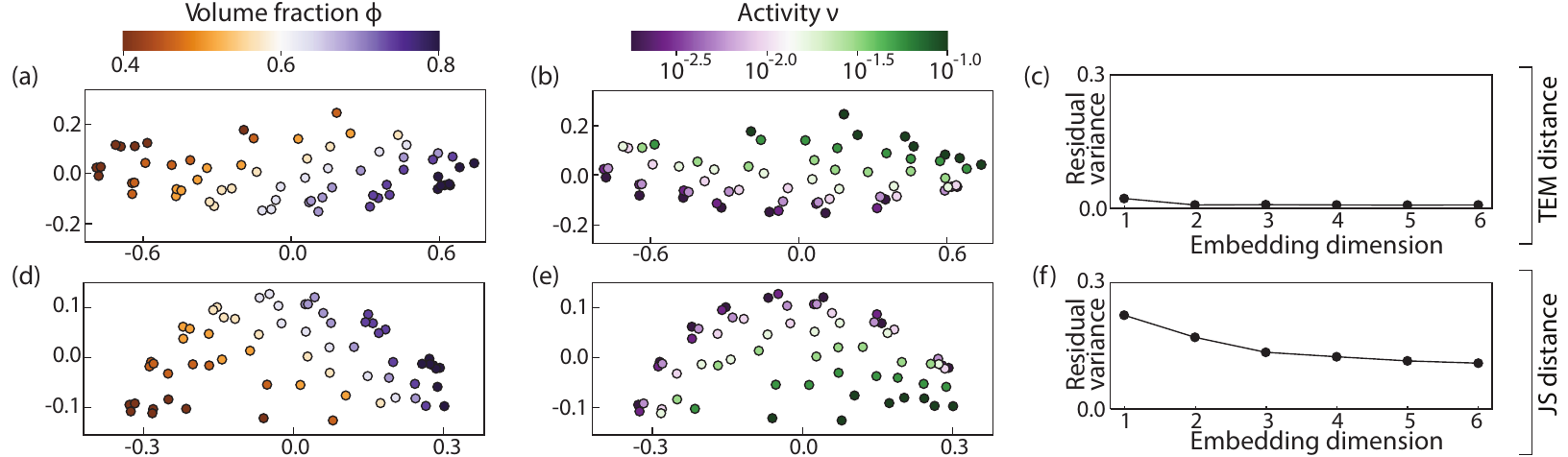}
\caption{\label{fig:JSABP}While both the JS and TEM distances recover the underlying phase space in a 2D MDS embedding of ABP simulations,
	only the TEM distance matrix can be comfortably embedded in a low dimensional space. (a-b) 2D MDS embedding of ABP simulations
	using the TEM distance, colored according to volume fraction and activity respectively. (c) over 98\% of the variance is recovered
	in a 2D MDS embedding of the TEM distance matrix. (d-e) 2D MDS embedding of ABP simulations with the JS distance. (f) even a
	6D MDS embedding only accounts for around 90\% of the variance of the JS matrix, with no clear
	preference for an embedding dimension.}
\end{figure}

For dense enough phase spaces, both the JS and TEM distances should be able to predict which points
are neighbors, even if the JS distance cannot tell how far distant points are. For this reason it
is not surprising that both distances, when embedded in 2D using MDS, recover the phase space for
the ABP simulations (Fig.~\ref{fig:JSABP}a,b,d,e). 
However, while the residual variance for the TEM distance
clearly indicates that the phase space is 2D (Fig.~\ref{fig:JSABP}c), and a 2D embedding 
recovers over 98\% of the variance, the JS distance shows no clear preference for any dimension,
and even a 6D embedding recovers only $90\%$ of the variance (Fig.~\ref{fig:JSABP}f). 
 
\begin{figure}
\includegraphics{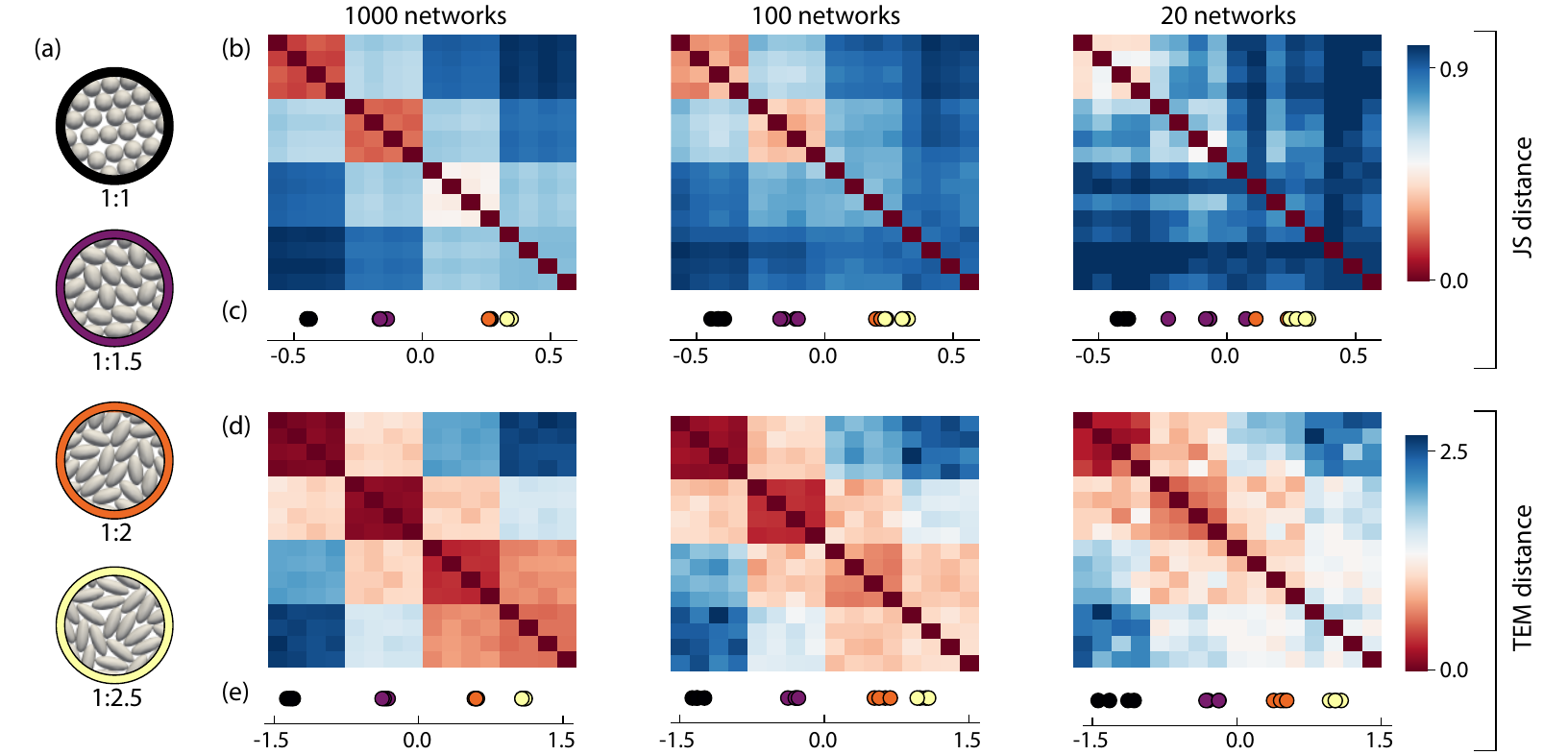}
\caption{\label{fig:Subsample}TEM distance can correctly distinguish ellipsoid packings from as little as
	20 local networks, unlike the JS distance. From a large jammed disorded ellipsoid packing simulation, 
	a small number (1000, 100, and 20) of local networks were sampled. This was done for aspect 
	ratios 1:1-2.5 shown in (a), with 5 simulations for each aspect ratio. (b) Distance matrix for 
	the JS distance. Taking 1000 networks it is apparent that there are 4 physically distinct materials,
	whereas for 20 networks this is not apparent. (c) 1D MDS embedding of the distance matrix in (b) with
	color coding according to (a). The correct MDS embedding is only recovered when 1000 
	networks are sampled. (d) Distance matrix for the TEM distance, even for only 20 sampled networks, 
	the 4 distinct parameter regimes are visible. (e) 1D MDS embedding of the distance matrix in (d) with
	color coding according to (a). The correct MDS embedding is recovered even when only 20 local networks
	are sampled.}
\end{figure}

The inability of the JS distance to tell how closely related networks are is a significant drawback when
limited data is available. In this regime, the true distribution is not sampled well; while this would
reduce the accuracy for all distances, a distance that has no concept of similarity between networks 
will particularly struggle. To see this, imagine the extreme case where for each distribution we take
only the one network that occurs most frequently. The TEM distance may still provide information about the
distributions, but the JS distance will give all distances as 0 or 1. To test this intuition, we took jammed disordered
packings of 10,000 ellipsoids for aspect ratios 1:1-2.5 (Fig.~\ref{fig:Subsample}a), 
and subsampled them, retaining either 1000, 100, or 20 
local networks. The TEM and JS distances were calculated (Fig.~\ref{fig:Subsample}b,d), as were their 
MDS embeddings, (Fig.~\ref{fig:Subsample}c,e). The correct MDS embedding is recovered for the TEM distance even
when only 20 local networks from each simulation are available, and this is consistent across simulations.
For the JS distance, the MDS embedding is incorrect even for 100 local networks. While the correct
ordering is recovered for 1000 local networks, there is hardly any separation between the 1:2 and 1:2.5
ellipsoids in the MDS embedding, unlike the 4 clear clusters that are apparent for all TEM MDS embeddings.
Therefore, in the case when limited data is available, the TEM distance outperforms the JS distance.
 
\subsection{Choice of radius $r$}

\begin{figure}
\includegraphics[width=\textwidth]{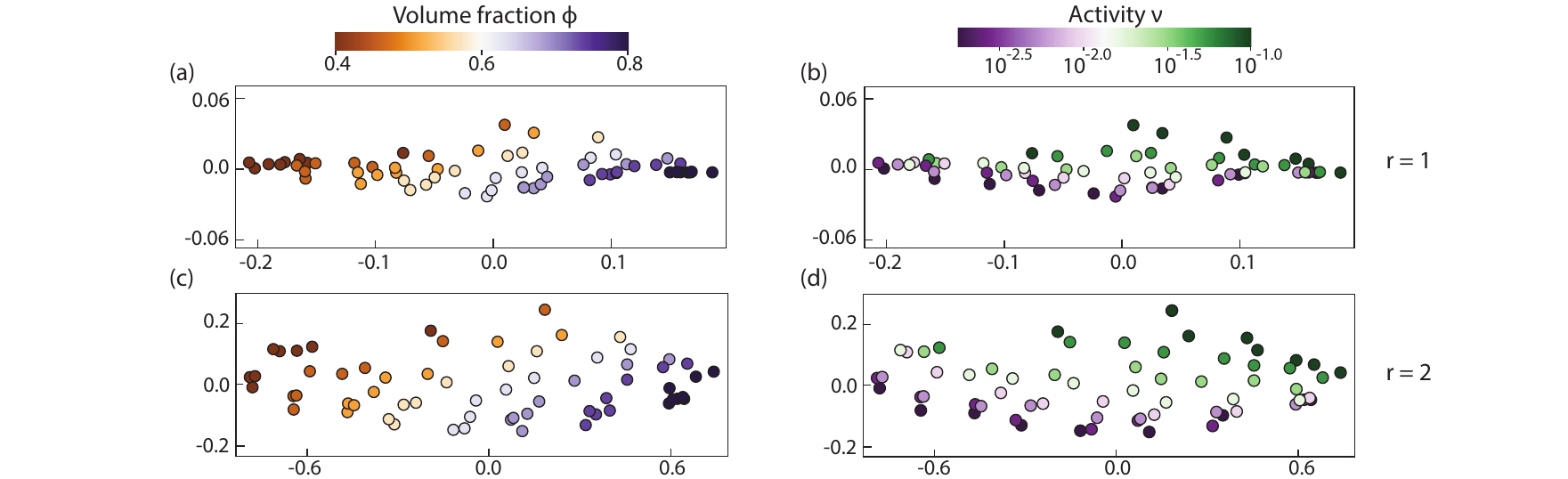}
\caption{\label{fig:r_comp}Taking the local network of radius $r=2$ is 
		sufficient to capture the 2D nature of the phase space, whereas 
		$r=1$ is insufficient. (a-d) MDS embedding of ABP simulations 
		for various values of volume fraction and activity. 
		(a-b) Taking local networks of radius $r=1$ recovers the volume 
		fraction as a principal component, (a), but fails to recover 
		the activity, except for intermediate values of the volume 
		fraction, (b).
		(c-d) Taking local networks of radius $r=2$ recovers a 2D phase 
		space with one principal component corresponding roughly to 
		volume fraction, (c), and the other to activity, (d).  
		}
\end{figure}

To capture the local ordering around a point we take the local network of radius $r$. The larger $r$ is, 
the more local information we capture, but also many  more distinct networks are observed
with corresponding increase in computational cost. Taking the ABP 
simulations as an example, 320 simulations of 2000 particles were performed, so 640,000 networks were
computed in total. For $r=1$, $O(20)$ distinct networks were observed, for $r=2$, $O(30,000)$ were observed and
for $r=3$, $O(500,000)$ were observed. This means that computing optimal transport exactly for $r \geq 3$ becomes computationally infeasible,
but could be solved approximately using entropic regularization~\cite{SMSolomonTransport}. However, since $r=2$ contains sufficient information
to recover the phase space, the question becomes whether $r=1$ may be sufficient as well. Although taking $r=1$ works to some extent,
it is not sufficient to recover the 2D phase space for the ABP simulations, Fig.~\ref{fig:r_comp}. Using the MDS embedding for $r=1$, over $99.6\%$ of the variance is in the first principal component (compared to $97.5\%$ for $r=2$), 
suggesting, erroneously, that the manifold is 1D, despite the true phase space lying on a 2D manifold.

\begin{figure}
\includegraphics{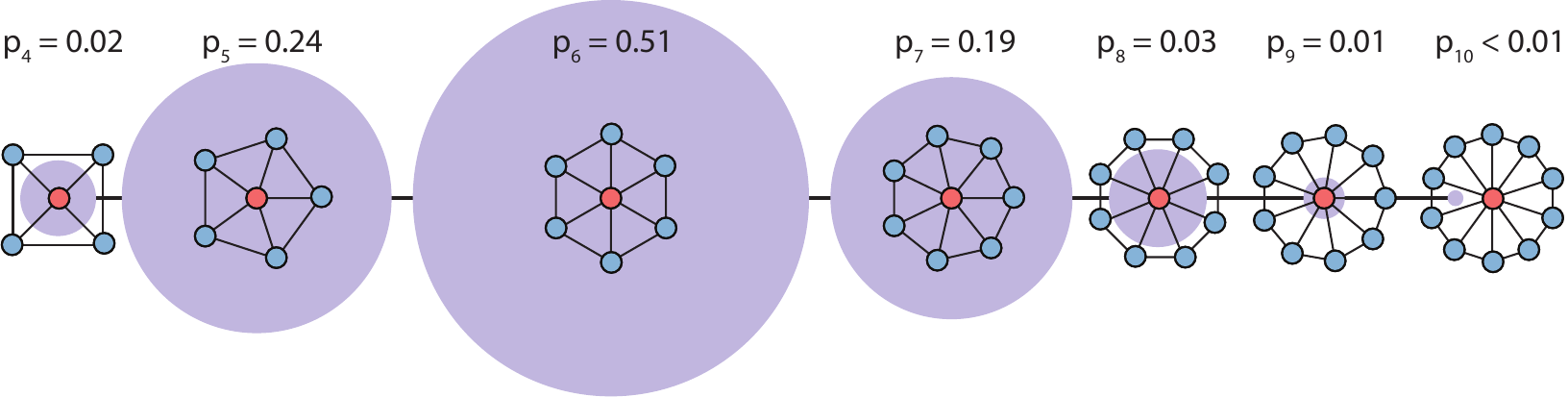}
\caption{\label{fig:FlipGraphr1}The flip graph for $r=1$ is essentially 1D
			and contains only the number of sides of each Voronoi cell. All
			ABP simulations were combined and the flip graph calculated for
			local networks of radius $r=1$. Each node of the flip graph is drawn
			in purple with the area of the node representing the proportion
			of local networks found in this state. The nodes that were not
			drawn represent less than $0.1\%$ of the total neworks seen.
			Each node has the local network that it represents overlayed. }
\end{figure}

To understand why taking $r=1$ is insufficient, we calculated the flip 
graph for all ABP simulations, and calculated the frequency at which 
each local network was observed (Fig.~\ref{fig:FlipGraphr1}). Neglecting
a negligible fraction of the total networks observed ($<0.1\%$), the flip
graph is simply a 1D, or path, graph and the networks tell us only how many
neighbors each Voronoi cell has. Euler's theorem tells us that the average
number of neighbors will be 6, and as demonstrated in 
Fig.~\ref{fig:FlipGraphr1}, the number of neighbors remains close to 6.
Therefore, the distribution is approximately 1D; if the fraction of
5 sided shapes is $p_5$, then the fraction of 7 sided shapes is 
$p_7 \approx p_5$, and the fraction of 6 sided shapes is $p_6\approx1- 2p_5$,
meaning the whole distribution is approximately described by a single parameter. While
in reality there are further degrees of freedom, the fact that the distribution is
almost 1D explains why $r=1$ will struggle to reconstruct a 2D or higher phase space.
\todo{
\subsection{Bacterial swarm}
The bacterial strain used for the swarming assay in this study was a $\Delta$\emph{comI} derivative of the 
ancestral \emph{B. Subtilis} strain NCBI3610, obtained from Daniel Kearns' lab at Indiana University. 
Cell cultures were grown at 37°C in Luria-Bertani (LB) liquid medium for 18h before incoluation. 
Swarming plates were prepared by filling $12ml$ of LB medium containing 0.5$\%$ bacto agar into a $90\text{mm}$ 
petri dish prior to a drying period of 20 minutes. The plates were then immediately inoculated with a 
small drop of bacterial culture of volume $<100nl$, which was placed in the center of the plate. 
Swarms were kept at 37°C in an enclosure containing a water reservoir and imaged for 12h using the 
automated microscopy routine and setup described in Ref.~\cite{SMJeckelPNAS}. At each space-time location
an image of $1024\times1024$ pixels is taken corresponding to an area of $170\mu\text{m} \times 170\mu\text{m}$.
}
\par
\todo{
To segment the cell centroids, we used the cell segmentation tool Stardist together
with a U-Net backbone~\cite{SMStardist,SMunet2015}. For training data, we
anually annotated 30 images containing 26,228 cells in total. The following table contains the hyperparameters, if not otherwise
stated the default parameters were used.}

\begin{center} \todo{
 \begin{tabular}{c c } 
 parameter & value \\
 \hline
 backbone & U-Net \\
 n\_rays & 32 \\
 unet\_n\_depth & 3 \\
batch\_size & 4 \\
patch\_size & (256,256) \\
\end{tabular}}
\end{center}

\todo{ 
There are 320 experimental snapshots across time and space which, due to the quadratic growth of computing the distance
matrix,  makes the full MDS embedding expensive (at least $16\times$ the
cost of the ABP embedding). The embedding by motif size parameters has no such quadratic cost, and was discussed in the main text.
Here we consider instead embedding all snapshots for a fixed time, and hence investigate the spatial structure at a particular time.
We performed this embedding for three time points, each containing 15-20 snapshots, Fig.~\ref{fig:SwarmMDS}. 
We first note that the swarm is heterogeneous, and the snapshots are separated by $O(500\mu\text{m})$ or hundreds of
bacteria lengths, meaning we do not necessarily expect the properties to change smoothly with radius. 
While there is often a relationship between the embedding
and the radius, the primary principal component does not always correspond to the radius (except for $t=8$h), Fig.~\ref{fig:SwarmMDS}. 
Volume fraction and speed are also strongly linked to the embedding and can explain certain principal components. Finally,
we find that the embeddings can cluster phases found by a data-driven clustering of snapshot parameters~\cite{SMJeckelPNAS}.
For the region of the swarm we are analyzing, there are three identified phases: rafting + biofilm precursor (R+BP), rafting (R), and 
single cell + rafting (SC+R)~\cite{SMJeckelPNAS}. We find that the phases tend to separate in the embedding, Fig.~\ref{fig:SwarmMDS}.
In addition to phase classification, we can now use the MDS embedding to see if a snapshot is truly representative of its
phase or if it is close to transitioning into a different phase.
\begin{figure}
\includegraphics{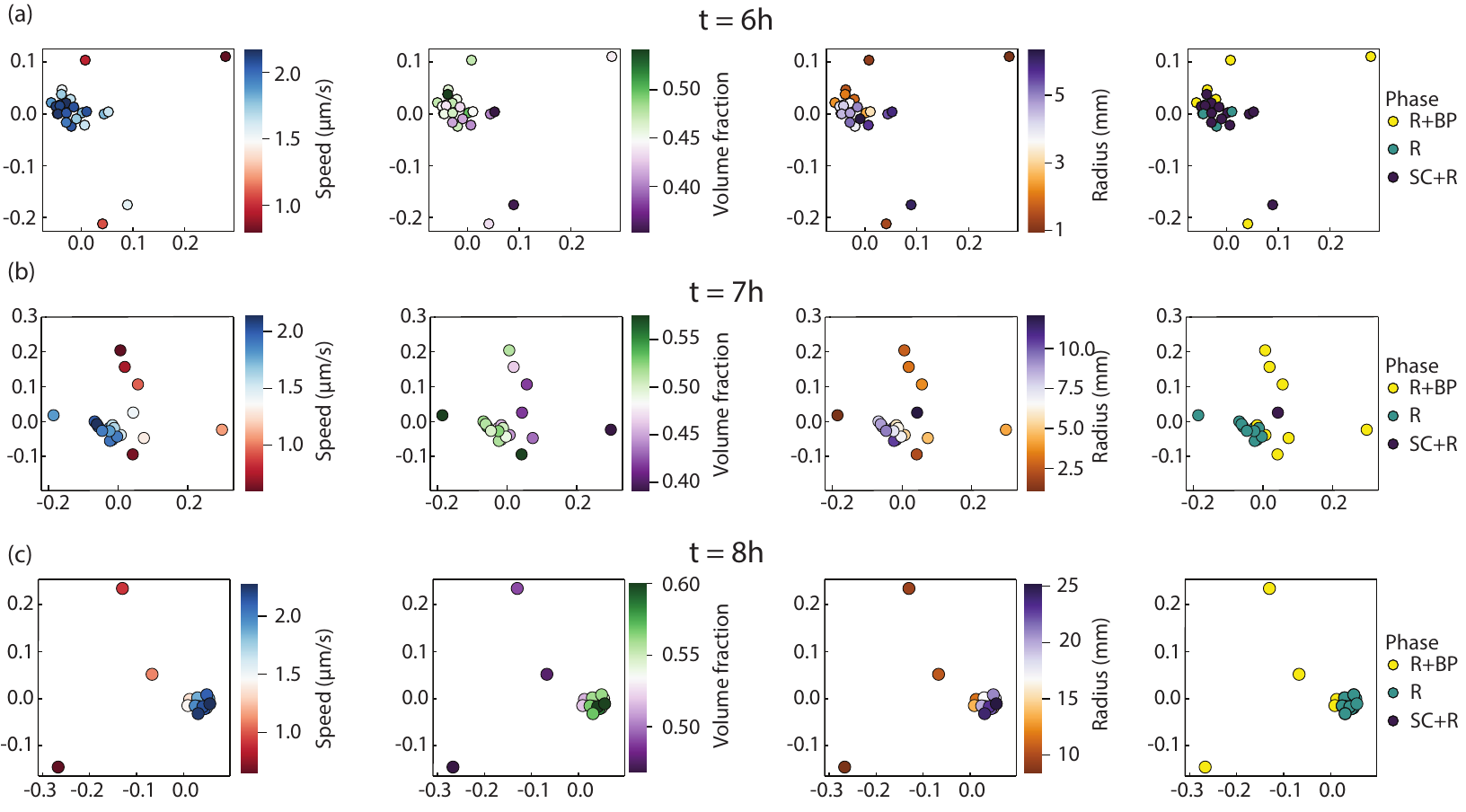}
\caption{\label{fig:SwarmMDS} \todo{MDS embedding of spatial snapshots at a fixed time find hereogeneous spatial structure. For times
$t = 6,7,8$h, all spatial snapshots were taken, pairwise distance matrix computed, and a MDS embedding was performed. The embeddings
are colored by speed, volume fraction, radius and phase. (a) At early times ($t=6$h), many snapshots contain fast moving cells and
this region of the embedding approximately recovers the spatial ordering. The remaining snapshots contain slower moving cells and
are either at the core of the swarm in the biofilm precursor region, or at the edge. (b) At mid times ($t=7$h), the spatial ordering
is not recovered, but either speed or volume fraction does a reasonable job at explaining the first principal component. The phases
are separated in this embedding, except some R+BP points which are close to the R cluster. (c) At late times ($t=8$h), either speed,
volume fraction, or radius could explain the first principal component. The phases are also separated.}}
\end{figure}
\subsection{Geodesics in $W_1$ space}
The energetic barrier required to perform a flip in epithelial cell layer suggests that the system might evolve in
such a way to minimize the total number of these flips required. To investigate, we seek to find a path between the average
start and end experimental distributions that exactly solves this minimization problem. Recall to find the earth mover's 
or $1^{st}$ Wasserstein ($W_1$) distance, we 
take the start and end distributions, say $P_0$, and $P_1$, and solve the for the optimal transport plan moving 
mass $\gamma_{ij}$ from $i$ to $j$. A natural way to interpolate this process would be to instead move mass
$t\gamma_{ij}$, from $i$ to $j$, for $t\in [0,1]$, which results in the intermediate distributions $P_t = (1-t)P_0 + tP_1$.
This path is a geodesic in the sense that $ \sum_{t_k} d(P_{t_k},P_{t_{k+1}}) = d(P_0,P_1)$, for 
$0 = t_0 \leq \cdots \leq t_{n} = 1$,  but $P_t$ is not the only interpolating
distribution to have this property, as geodesics are not unique under the $W_1$ metric~\cite{SMsolomon2016continuous}. 
This distribution could be interpreted as phase separated
growth, a fraction $(1-t)$ of the system is in the initial distribution, the remainder in the final, and the size of the
phase in the final distribution grows until the whole system is in that distribution. However, this has a certain 
unphysical aspect of mass instantaneously appearing in the final state as the path is traversed~\cite{SMsolomon2016continuous}. 
In contrast, it is well known that when computing the 
second Wasserstein ($W_2$) distance on a continuous manifold $\mathcal{M}$, one finds a naturally interpolating path
between initial and final distributions. Moreover, the distance can be computed as
\begin{align}
W_2^2(\rho_0,\rho_1) &= \min_{\rho,v} \int_{t=0}^{t=1}\int_{\mathcal{M}} \rho(x,t) || v(x,t)||^2\, \mathrm{d}x \, \mathrm{d}t\\ \nonumber
&  \;\; s.t. \; \rho(x,0) = \rho_0, \; \rho(x,1) = \rho_1, \text{ and} \\ \nonumber
& \quad \frac{\partial \rho}{\partial t} + \nabla \cdot (\rho \textbf{v}) = 0.
\end{align}
for a probability density $\rho$ over $\mathcal{M}$ advected by a velocity field, $v$~\cite{SMsolomon2016continuous}. 
The term in the integrand to be minimized is the natural analogue of dissipation for this transport problem, and
the intermediate path $\rho(x,t)$ is the natural interpolation between start and end 
distributions~\cite{SMsolomon2016continuous}.
}

\todo{
In order to compute a $W_1$ geodesic on a graph, we must find a transport
plan $\gamma_{ij}(t)$, that minimizes
\begin{equation}
W_1 = \sum_{ij} \int_t \gamma_{ij}(t) \mathrm{d} t
\end{equation}
together with the conservation equation
\begin{equation}
\frac{d}{dt} P_t(i) = \sum_{j} \gamma_{ji}(t) - \sum_{j} \gamma_{ij}(t),  
\end{equation}
with initial and final conditions $P_{t=0} = P_0$ and $P_{t=1} = P_1$; $\gamma_{ij}(t) \mathrm{d} t$ is the mass transported
from $i$ to $j$ in time $\mathrm{d}t$.
As previously discussed, there are many such $\gamma_{ij}(t)$ that satisfy this 
equation. Here, we take all transport plans $\gamma_{ij}(t)$ that minimize the $W_1$ cost, and out of these, find the plan
that minimizes the transport dissipation. To write down the graph-based equivalent of the transport dissipation, we 
can interpret the mass transport $\gamma_{ij}(t) = (\text{density}) \times (\text{velocity})$, so that dissipation would
become,
\begin{equation}
R = \sum_{i,j} \int_t \frac{\gamma_{ij}(t)^2}{P_t(i,j)} \mathrm{d} t,
\end{equation}
where the sum is assumed to be taken over non-zero $\gamma_{ij}$, and $P_t(i,j)$ denotes the 
probability density over edge $(i,j)$. To define an edge based density (previously density was defined on vertices),
we follow~\cite{SMsolomon2016continuous}, taking the harmonic mean of the densities at the corresponding vertices,
$P_t(i,j) = 2P_t(i)P_t(j) / (P_t(i) + P_t(j))$.
To solve numerically, we discretize in time, finding $T+1$ states, $P^{(0)}, P^{(1)}, \dots, P^{(T)}$, and
$T$ transport plans $\gamma^{(1)}, \dots \gamma^{(T)}$, making the transport dissipation~\cite{SMsolomon2016continuous},
\begin{equation}
R =  T \sum_{l=1}^{l=T} \sum_{i,j} (\gamma_{ij}^{(l)})^2\left( \frac{P^{(l)}(j) 
+ P^{(l-1)}(i)}{2P^{(l)}(j) P^{(l-1)}(i)} \right),
\end{equation}
where the sum over $i,j$ is assumed to only include $\gamma_{ij}^{(l)} > 0$ terms so that $P^{(l)}(j)$, $P^{(l-1)}(i)$ are
always non-zero when they appear in the sum. Note that the advective nature of transport is reflected in the harmonic mean by
taking the density at the source of the transport at time $l-1$ and the density at the destination of transport at time $l$.
The full discretized problem for finding the dissipation regularized geodesic
can be stated as
\begin{align}
\min_{P^{(l)}, \gamma^{(l)}} R(P^{(l)},\gamma^{(l)}) \;\; & s.t.  \;\; P^{(0)} = P_0, P^{(T)} = P_1, \\ \nonumber
& \sum_j \gamma_{ji}^{(l)}
- \sum_j \gamma_{ij}^{(l)} = P^{(l)}(i) - P^{(l-1)}(i), \\\nonumber
& \sum_{l=1}^{T} \sum_{ij} \gamma_{ij}^{(l)} = W_1(P_0,P_1),
\end{align}
where the last constraint enforces that the path is still optimal with respect
to the $W_1$ metric. The integer parameter $T$ determines the level of discritization, with the continuous limit being recovered
for $T\to \infty$~\cite{SMsolomon2016continuous}. This problem can be solved by second order conic optimization programs, and whilst significantly slower than 
solving the linear min-cost flow for $W_1$, it is still practical for the systems considered here.}

\todo{
We see that the dissipation regularized geodesics calculated 
by this method closely match the actual path the data takes, unlike the phase separated interpolation, Fig.~\ref{fig:Geodesics}.
Fluctuations around the geodesic mean that the system will not exactly minimize dissipation, but nevertheless follows
the minimizing path closely.
\begin{figure}
\includegraphics{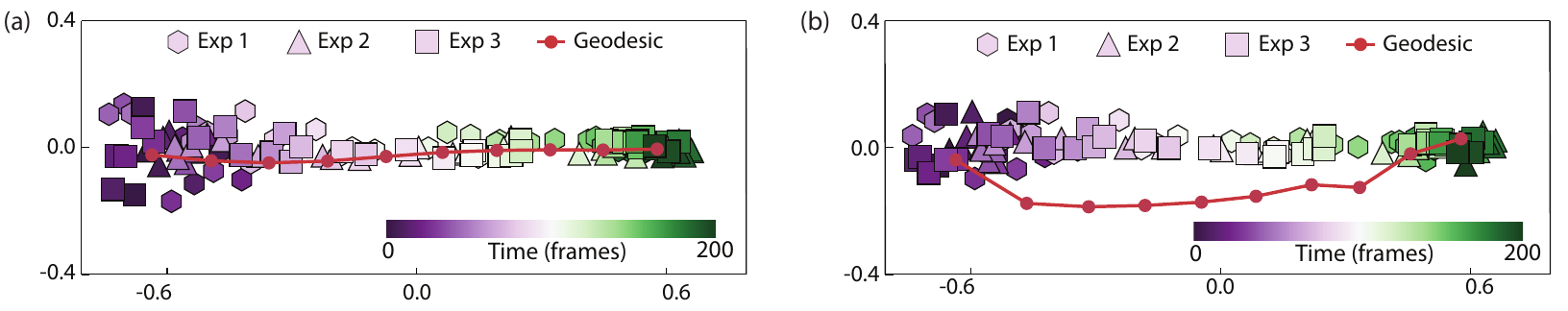}
\caption{\label{fig:Geodesics} \todo{Comparison of dissipation regularized geodesic and phase separated growth geodesic. (a) MDS
embedding of data and dissipation regularized geodesic (also Fig. 3h of main text). (b) MDS embedding of data and the phase
separated growth geodesic.}}
\end{figure}
\subsection{Topological entropy}
In practice, if the underlying parameters are unknown, interpretation
of the principal components directly from the data is desirable. In the next section, we introduce tools to 
analyze the motif size 
distribution, which yields further interpretations of the embedding.
Here, we note that from the observed motif distributions $P(i)$, we can compute the topological Shannon entropy~\cite{SMEntropyLazar}, 
$S= -\sum_i P(i) \ln P(i)$. The interpretation of entropy is that it is a measure of how many
states are available to the system.
The entropy $S$ often explains the first MDS principal component, for instance in the ABP embedding, 
Fig.~\ref{fig:Entropy}. In the \textit{Drosophila} example, the entropy corresponds to the first principal component, 
notably decreasing with time, Fig.~\ref{fig:Entropy}, consistent with known results about the increasing regularity
of the developing wing~\cite{SMDrosophilaExp}. \begin{figure}
\includegraphics{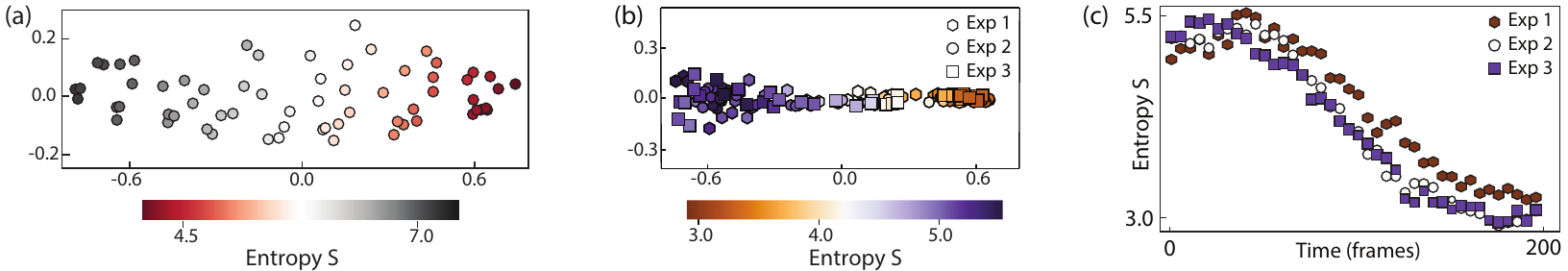}
\caption{\label{fig:Entropy}\todo{Topological entropy as an analysis tool for embeddings. (a) MDS embedding of ABP data colored by
topological entropy shows that the topological entropy corresponds to the first principal component, and increases with decreasing
density as more states become available. (b) MDS embedding of \textit{Drosophila} data colored by topological entropy. Topological entropy 
corresponds to the first principal component. (c) Topological entropy decreases with time for the \textit{Drosophila} embryos as the cell layer
becomes more regular.}}
\end{figure}
\subsection{Tracy-Widom distribution}
We seek to understand the structure of the data by understanding the probability distribution of motifs. 
This space is hard to analyze directly, so we consider a projection onto
$\mathbb{N}$, by taking the distribution of motif sizes, defining the size of a motif to be the number of vertices that it has.
In the main text, we claimed that for the swarm data, the 
distributions differed in mean and variance, but all lay on a universal distribution, namely
the Tracy-Widom (TW) distribution. Here we provide stronger numerical evidence that the distribution is indeed TW by 
analyzing the network formed by a Poisson point process, known as a Poisson-Voronoi (PV) 
tesselation~\cite{SMLazarPV}, which is equivalent to a non-interacting gas, taking $2\times10^6$ points in a periodic domain. To analyze the motif size distribution, we took the 
observed distribution for fixed $r$, and rescaled so that it has mean 0, variance 1.
We compare this against various known distributions including a Gaussian, Log-normal, and TW for all $\beta$ parameters, 
all rescaled to have unit variance and zero mean. 
The distribution were chosen for their universal characteristics, and hence plausible appearance in the motif size 
distribution. Gaussian distributions are universal due to the central limit theorem, and for positive variables
a Log-normal distribution has universal properties~\cite{SMlimpert2001log}. We discuss in the main text the universal character of TW
distributions. The parameters for the Log-normal were chosen by maximum likelihood estimation before rescaling, the TW and Gaussian
distributions were only rescaled, not fitted. We see for $r=2$, that the distribution lies enough close to TW to be plausible,
but differs slightly in the left tail and due to the discrete nature of motif sizes, does not finely resolve the distribution, 
Fig.~\ref{fig:PoissonVoronoi}a. However, by taking ever larger motif sizes, $r=3,4$, the observed distribution gets significantly
closer to TW, matching in both the $O(1)$ regions and in the tails until at least probability density $10^{-4}$, 
Fig.~\ref{fig:PoissonVoronoi}b-c. In particular, the distribution for larger motif sizes lies on TW, and not a Gaussian or 
Log-normal, even though Log-normal (for our parameters) and TW are very similar distributions. 
TW with $\beta=2$ appears to be the best fitting
of the TW distributions, although they only differ slightly when rescaled, so further numerical tests would be needed to confirm this.
In the main text, and the rest of this document we use $\beta=2$ for the TW distribution. We note that this distribution is 
universal in the sense that the motif size distributions for PV and for the swarm data do not share the same mean or variance, and
the underlying processes are different. The mean $r=2$ motif size of a PV process is 20.7 and variance 14.8, which is
significantly greater than the corresponding quantities measured from the swarm snapshots. Further, PV can be thought of as a snapshot of an non-interacting ideal gas,
whereas the bacteria in the swarm have strong excluded volume interactions and density plays an important role. Nevertheless, the
same distribution is found. 
\begin{figure}
\includegraphics{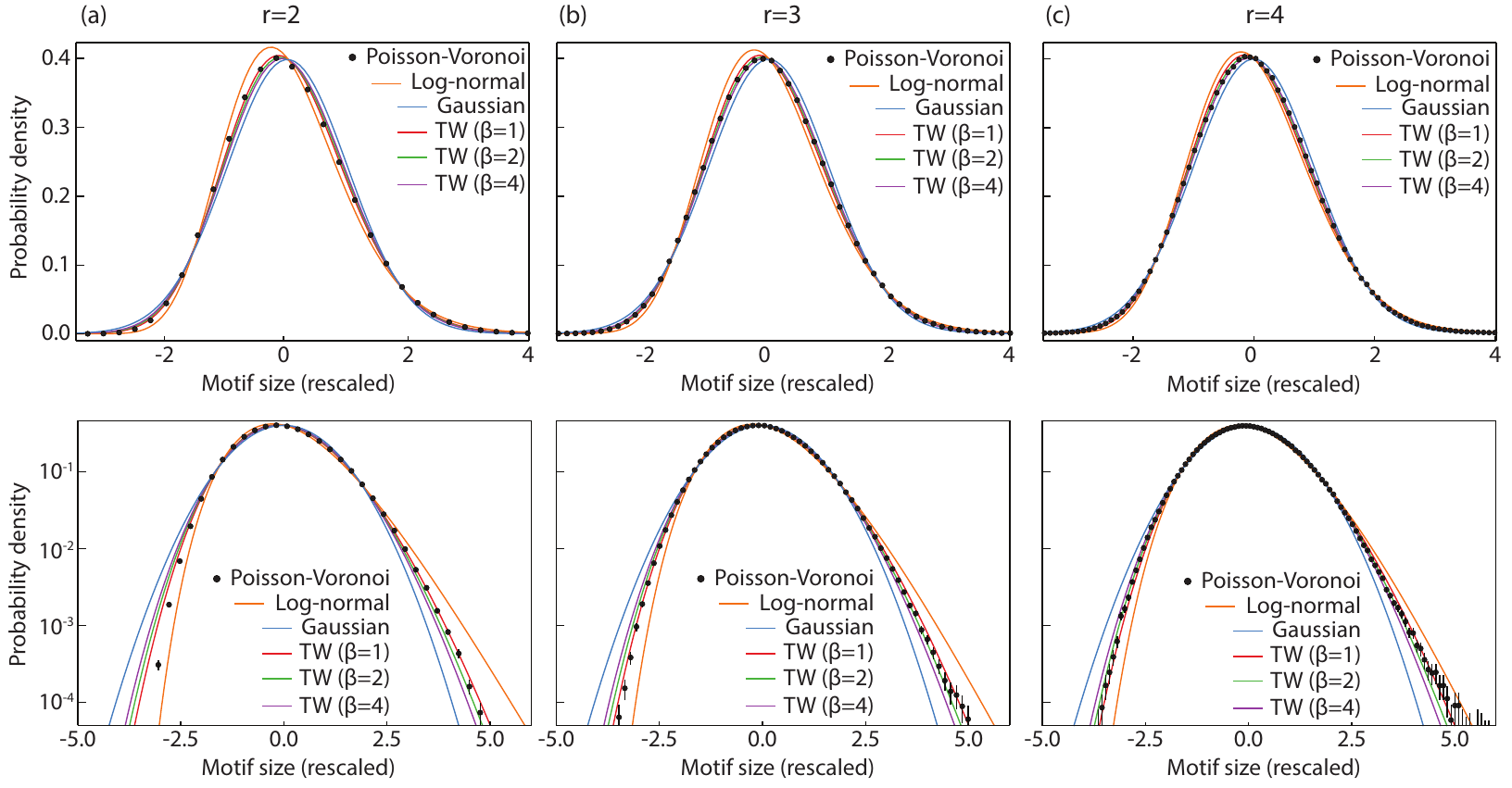}
\caption{\label{fig:PoissonVoronoi} \todo{Rescaled motif size distribution for various motif radii, compared against known distributions.
(a) For $r=2$, the distribution is close to TW, but differs in the left tail due to geometric constraints on the minimum motif 
sizes. (b) For $r=3$, the distribution approaches TW on the left tail, but does not yet lie exactly on it. (c) For $r=4$, 
the distribution closely matches TW in the regions shown. Error bars are $\pm2$ standard deviation, from sampling error.}}
\end{figure}
\subsection{Analysis of motif size distribution}
Applying the motif size analysis to the ABP data, we find a range of possible distributions, Fig.~\ref{fig:ABPDist}d,
not all of which appear to be TW. We can approximately characterize these distributions by taking their mean,
variance, and third moment. We find that the mean and variance correspond to the first principal component of the MDS
embedding, and the third moment corresponds to the second principal component, Fig.~\ref{fig:ABPDist}(a-c). In an alternative
to the TEM-MDS embedding, one can plot the simulations in the variance-third moment plane, finding that this again recovers the
activity/volume fraction phase space, Fig.~\ref{fig:ABPDist}(e-f). 
This is not preferable to the metric embedding, since there is no notion of distance, 
although here one need not compute $O(N^2)$ components of a distance matrix, providing an 
alternative when $N$ is large. To understand these embeddings, we see that reducing the volume
fraction for fixed activity causes more states to become available as particles interact less. Hence the motif size variance
increases and so does the third moment (as does the topological entropy). Interestingly, when the activity is increased, 
the variance does not change much, but the third moment increases significantly. In the most extreme case of $\phi = 0.8$, 
going from the lowest activity to the highest, the third moment increases by a factor of 16, whereas the variance increases
by a factor of 1.2, and the mean does not chance. Hence activity changes the distribution
by increasing the probability of large motif fluctuations, 
which directly correspond to the giant number fluctuations of the phase separated regime~\cite{SMMarchetti2014}. 
\\
\begin{figure}
\includegraphics{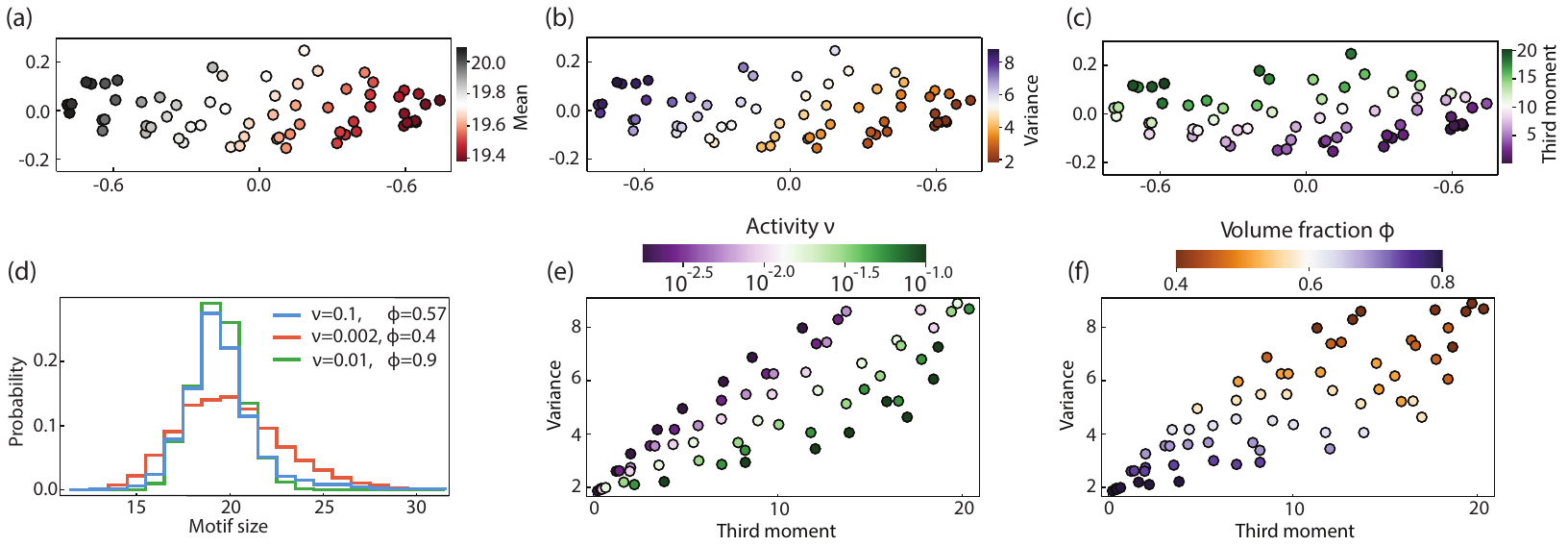}
\caption{\label{fig:ABPDist}\todo{Motif size distributions characterize ABP simulations. (a-c) Coloring the MDS embeddings by mean,
variance and third moment respectively, finds that the mean and variance correspond to the first principal component, and the
third moment corresponds to the second principal component. (d) Example motif size distributions for various values of
volume fraction and activity. (e-f) Simulations embedded in the variance-third moment plane colored by activity and volume 
fraction respectively. The embedding, like the MDS embedding, recovers the parameter space.}}
\end{figure}
\begin{figure}
\includegraphics{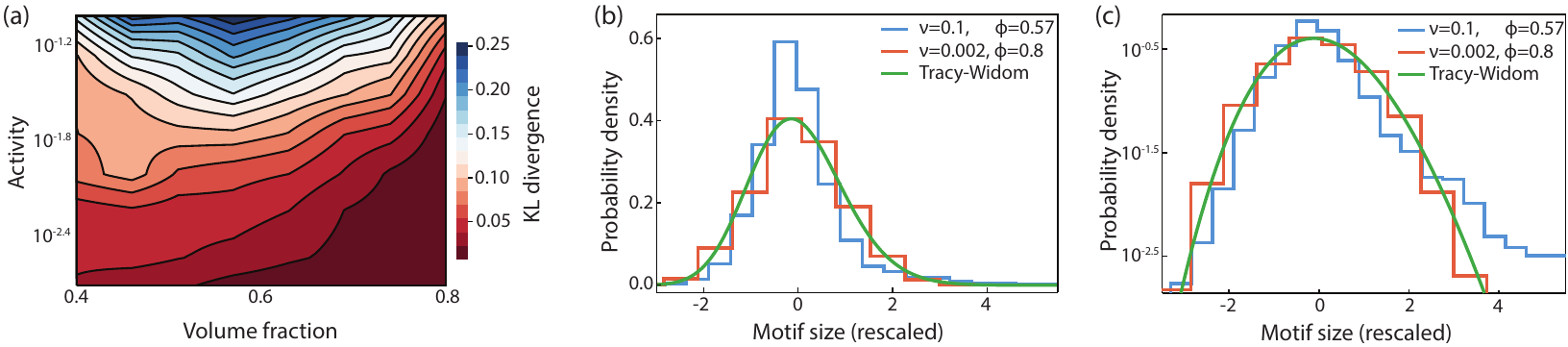}
\caption{\label{fig:ABPTW}\todo{Motif size distributions for ABP simulations are approximately Tracy-Widom in the liquid-like region of phase space.
		(a) KL divergence between rescaled motif size distributions and TW distributions shows regions of phase
		space where the TW approximation holds. (b) Representative example histograms of distributions which are far from TW, 
		and distributions that are close to TW, together with the TW distribution. (c) Same as (b) with log-scale.}}
\end{figure}
The distributions seen in the ABP simulations appear to take a variety of forms, although the Tracy-Widom (TW) distribution appears,
Fig.~\ref{fig:ABPTW}. We evaluate where in the parameter space the motif distribution appears most TW like, by comparing
the distribution rescaled to zero mean and variance 1, and the TW distribution similarly scaled. We use the Kullback-Leibler (KL)
divergence between two distributions, defined as $KL(P||Q) = \sum_n P(n) \log( P(n)/Q(n) )$, representing the amount of 
information lost when the data distribution ($P$) is approximated by the TW distribution ($Q$).
To compare continuous and discrete distributions in calculating the KL divergence, we consider the 
discritized TW pdf to be the mass lying within a bin as defined by the histogram of the discrete data. 
We find that in the most active
regimes the observed distribution is furthest from TW. Perhaps this is because in the phase separated regimes, the dense phase
is a strongly coupled glass like phase, with a narrow distribution of motif sizes, but the gas-like phases can take a wide variety
of motif sizes, leading to unusual distribution tails. The region where the distribution is closest to TW, is the region
where the system is most liquid like.}

\begin{figure}[h] 
\includegraphics{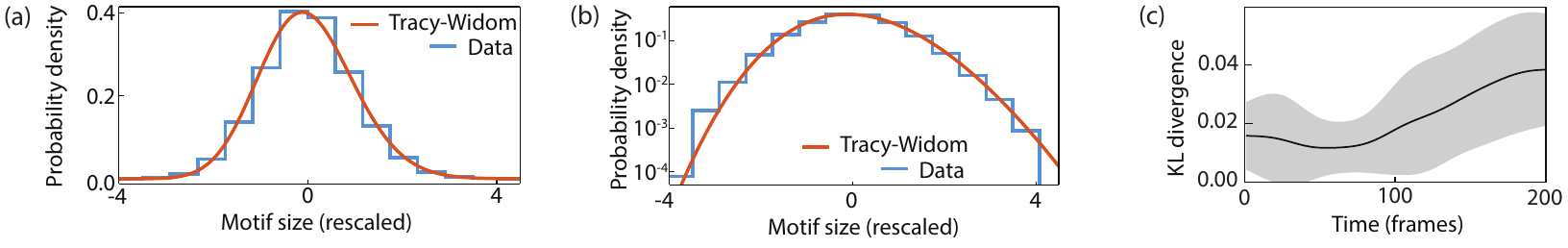}
\caption{\label{fig:DrosophilaTW}\todo{Motif size distribution for the wings of \textit{Drosophila} embryos is well approximated by TW
			at early times, but takes a different form at later times. (a) We combinine the first 100 frames into a single
			distribution, and rescale to have zero mean and variance 1. The resulting histogram is well approximated by 
			the TW distribution.
			(b) same as (a) with log scale. (c) The KL divergence between the observed distribution and TW remains low until
			around 100 frames where it increases. Black line is local average, grey region shows $\pm 2\times$(standard deviation).}
}
\end{figure}

\todo{
For the \textit{Drosophila} data, we find that the motif size distribution is well approximated by TW in early times, whilst the
EC layer is irregular and the system is most liquid-like. However, at later times when the final state of a more regular lattice
appears, the distribution is no longer close to TW, Fig.~\ref{fig:DrosophilaTW}.}

\todo{
For the bacterial swarm data, while the motif distributions have varying mean and variance, 
the distributions appeared to have a common form when rescaled to have 0 mean and variance 1, Fig.~\ref{fig:SwarmTW}a.
We rescale every observed motif by the mean and variance of the distribution at that snapshot
and combine in a single histogram. Since all distributions have similar mean, and are originally defined on the integers,
the combined histogram still shows discrete peaks, Fig.~\ref{fig:SwarmTW}b. We therefore choose histogram bins so that
each contains exactly one discrete spike, and plot them as points so that the height represents the probability
density, the $x$ location represents the center of mass, with horizontal bars of $\pm(\text{within bin standard deviation})$ 
representing the width of the peak. We compare this against various known distributions, Fig.~\ref{fig:SwarmTW}(c-d), as before
with the PV tesselation. We once again find that TW fits the data well, although since we are using $r=2$ here, they disagree
slightly at the left tail.
\begin{figure}
\includegraphics{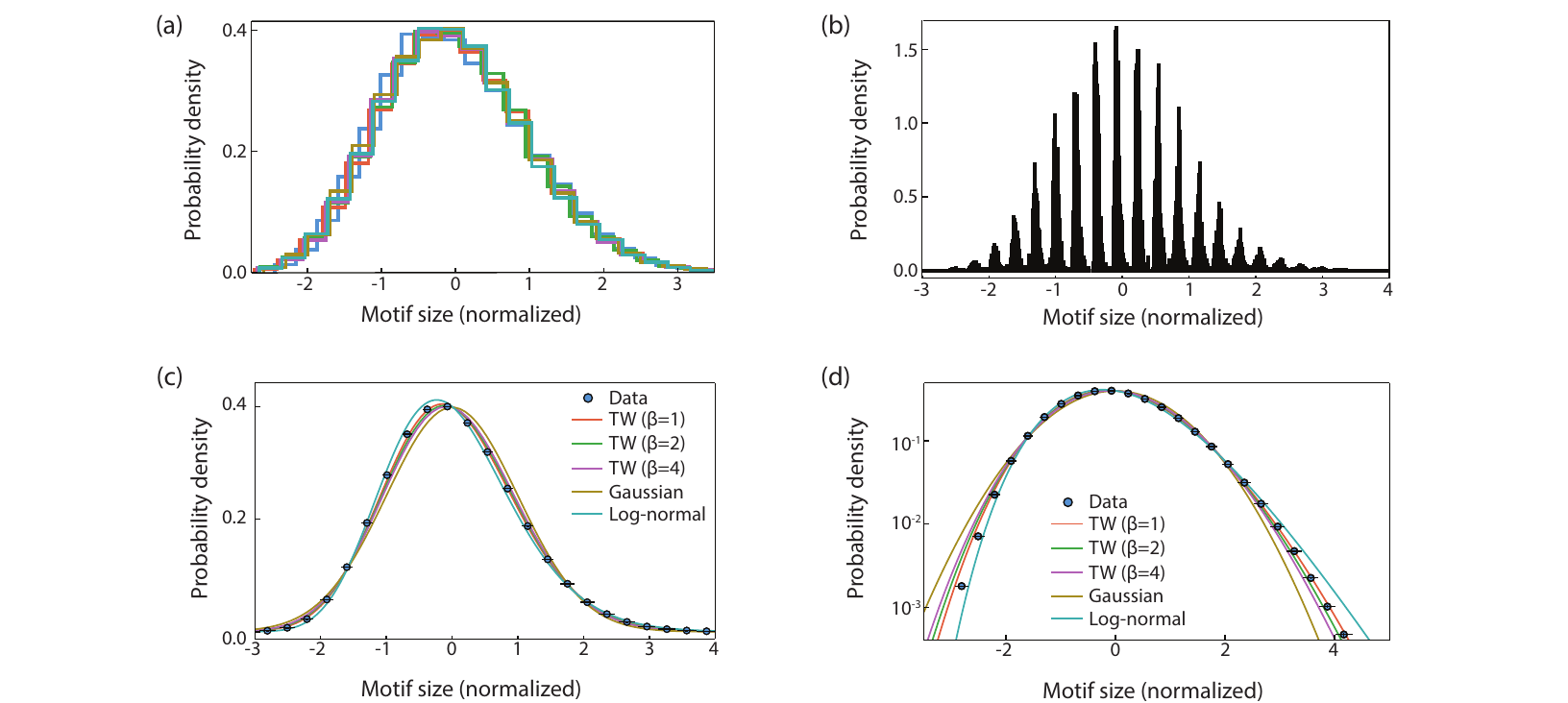}
\caption{\label{fig:SwarmTW}\todo{Observed motif size distribution against plausible known distributions. (a) Randomly chosen
snapshot distributions rescaled lie on the same curve. (b) Overall motif distribution with each snapshot data rescaled to have
0 mean and variance 1. Peaks are visible due to the integer sizes of the motifs, as well as a smooth envelope. (c-d) Histogram
with bins containing a single peak, compared against known distributions with normal (c) and log scale (d).}}
\end{figure}
}

\end{document}